\documentclass[11pt]{article}

\usepackage{setspace}
\usepackage[margin=1.5in]{geometry}

\usepackage[english]{babel}
\usepackage[utf8x]{inputenc}

\usepackage{amsmath}
\usepackage{amsfonts}
\usepackage{amssymb}
\usepackage{graphicx}

\numberwithin{equation}{section}

    \usepackage{tikz-cd}
    \usepackage{mathrsfs}
    
    \usepackage{url}
    \usepackage{color}
    \usepackage[colorlinks=true, citecolor=blue]{hyperref}
    \date{\today}

\def\a{\alpha}   
\def\b{\beta}   
\def\beq{\begin{equation}}
\def\eeq{\end{equation}}
\def\ra{\rangle}
\def\la{\langle}

\newcommand{\ca}[1]{\mathcal{#1}}
\newcommand{\bb}[1]{\mathbb{#1}}
\newcommand{\al}[1]{\mathfrak{#1}}
\newcommand{\scr}[1]{\mathscr{#1}}
\newcommand{\ii}{\imath}

\title{\bf Particle on the sphere:\\ group-theoretic quantization\\ in the presence of a magnetic monopole
\vspace{5mm}
}
\date{}
\author{Rodrigo Andrade e Silva\thanks{rasilva@umd.edu}
~~and~~
Ted Jacobson\thanks{jacobson@umd.edu}
\vspace{-5mm}
\\
\and
{\it \small Center for Fundamental Physics,
University of Maryland}\\
{\it\small College Park, MD 20742, USA}}

\begin{document}
\begin{titlepage}
\maketitle
\thispagestyle{empty}

\begin{abstract}
The problem of quantizing a particle on a 2-sphere has been treated by numerous approaches, including Isham's global method based on unitary representations of 
a symplectic symmetry group that acts transitively on the phase space. Here we reconsider this simple model using Isham's scheme, enriched by a magnetic flux through the sphere via a modification of the symplectic form. 
To maintain complete generality we construct the Hilbert space directly from the symmetry algebra, which is manifestly gauge-invariant, using ladder operators. 
In this way, we recover algebraically the complete classification of quantizations, and the corresponding energy spectra for the particle.
The famous Dirac quantization condition for the monopole charge follows from the requirement that the classical and quantum Casimir invariants match.
In an appendix we explain
the relation between this approach and the more common one that assumes from 
the outset a Hilbert space of wave functions that are sections of a nontrivial line bundle over the sphere, and
show how the Casimir invariants 
of the algebra determine
the bundle topology. 
\end{abstract}

\end{titlepage}

\tableofcontents

\section{Introduction}
\label{sec:intro}

Canonical quantization is a magic wand, discovered by Dirac, that transmogrifies a classical dynamical theory into a corresponding quantum theory, often in perfect agreement with observations. However, for most classical theories
Dirac's procedure depends on the choice of phase space coordinates over which to wave the wand, so the resulting quantum theory is ambiguous. Moreover, a generic phase space has nontrivial topology, and does not even admit a global coordinate chart. In complete generality, the only recourse is to accept the ambiguity, and to explore all quantizations.
But some classical dynamical systems
possess symmetries that can be used to identify a restricted class of quantizations which preserve these symmetries in the quantum theory. Such quantizations would obviously yield the best guess, if indeed the original classical theory is the classical limit of some quantum theory. 

Isham provided a generalization of Dirac's canonical quantization that is designed to preserve a chosen transitive group of phase space symmetries, and can be applied to topologically 
nontrivial phase spaces~\cite{isham1984topological, isham1989canonical}. 
Our primary 
interest in 
this, 
as was Isham's, 
is ultimately to restrict the possibilities for nonperturbative quantization of general relativity. 
But to develop understanding of how the scheme works, the ambiguities that remain, and the relation to other quantization schemes, it is useful to consider simpler systems. A particle on a 2-sphere is one of the simplest such systems, and has already been treated by many different approaches, including Isham's~\cite{isham1984topological, isham1989canonical, kleinert1990path, kleinert1997proper, landsman1991geometry,ohnuki1993fundamental,mcmullan1995emergence, dictua1997quantization,  woodhouse1997geometric,neves2000stuckelberg, bouketir2000group, kowalski2000quantum, hong2000improved,
abdalla2001quantisation,hong2004gauged,
liu2011geometric, hall2012coherent, kemp2014geometric, zhong2015enlarged,ouvry2019anyons}. 
Here we 
consider this simple model, enriched by the inclusion of a magnetic flux through the sphere, 
with the aim of implementing
Isham's quantization scheme without making any choices other than that of the group of canonical symmetries.\footnote{A similar approach appears to have been considered in \cite{bouketir2000group}, however we could not obtain access to the relevant part of the document. Related work is also mentioned in \cite{bouketir1999quantization} by the same author, but it refers to a preprint that we could not find.}  In order to maintain complete generality for the  unitary representations of the quantized algebra,  we construct the Hilbert space directly from the algebra, rather than adopting the framework of wave functions. In this way, we recover algebraically the complete classification of quantizations, as
well as the famous Dirac quantization condition for the monopole charge and the corresponding energy spectra.
We also explain the relation between this approach and the more common one that assumes from 
the outset a Hilbert space of generalized wave functions that are sections of a nontrivial line bundle over the sphere. 

Isham applied his quantization scheme to the phase space $\ca P = T^*S^2$, the cotangent bundle of the 2-sphere, with the canonical symplectic form $\omega =
dp_i\wedge dq^i$.  
He found that the Hilbert space must carry some unitary irreducible (projective) representation of the 3-dimensional Euclidean group, $E_3 = \bb R^3 \rtimes SO(3)$. From Mackey's theory of induced representations, one obtains that each irreducible unitary representation $\scr U^{\!(n)}$ of $U(1)$, labeled by an integer $n$, yields a Hilbert space represented by sections of a certain bundle, $\scr U^{\!(n)}$-associated to the Hopf bundle $SU(2) \rightarrow SU(2)/U(1)\sim S^2$. (See Appendix \ref{app:Bundle} for details.)
These sections can be seen as ``twisted'' wavefunctions over the sphere, with the ``twisting'' described by the first Chern number of the bundle (which is $n$ for the $\scr U^{\!(n)}$-associated bundle). The choice of this integer $n$ is equivalent to the assignment of an intrinsic spin to the particle, and its value is fixed to be zero if one
imposes, as a correspondence principle, that classical Casimir invariants of the Poisson algebra are preserved upon quantization. When the particle is electrically charged, and a magnetic monopole field is included, we find again the same classification of Hilbert spaces, but the correspondence principle for Casimir invariants determines 
$n$ in terms of the product of the monopole charge with the electric charge. The inclusion of the magnetic monopole field is thus equivalent to the assignment of an intrinsic spin to to the particle.

The usual prescription to include coupling to a magnetic field 
is to make the replacement $p \rightarrow p - eA$ in the Hamiltonian, where $p$ is the momentum, 
$e$ is the electric charge of the particle and $A$ is the magnetic potential 1-form in some local gauge. 
If $A$ is defined globally on the sphere then 
the magnetic field $B=dA$ is an exact 2-form, so the net magnetic flux 
through the sphere must vanish.
In the presence of a nonzero net magnetic flux, therefore, 
$A$ cannot be defined globally on the sphere, and hence the Hamiltonian is not globally defined.
This can be accommodated by defining the Hamiltonian in local gauge patches, and accompanying a 
gauge transformation $A\rightarrow A + d\lambda$ with a corresponding symplectic (canonical) transformation 
$p\rightarrow p + ed\lambda$. In this way, however, the canonical momentum ceases to be an observable, and 
a global description is lacking.

Instead, we shall maintain manifest gauge invariance and a globally defined Hamiltonian 
by incorporating the magnetic field into the symplectic structure~\cite{sniatycki1974prequantization, sternberg1977minimal}.   
That is, we replace
$p$ by $p + eA$ in the symplectic form $dp_i \wedge dq^i $. This results in 
$d(p_i + eA_i) \wedge dq^i = d(p_i dq^i) + e d(A_i dq^i)$, which can be written covariantly as 
\begin{equation}\label{sym}
\omega = d\theta + e \pi^* B \,.
\end{equation}
Here $\theta$ is the canonical symplectic potential, defined by
$\theta(X) = p(\pi_* X)$, where $X$
is a tangent vector on the cotangent bundle $T^* S^2$,
and $\pi: T^* S^2 \rightarrow S^2$ 
is the bundle projection map
(with $\pi^*$ and $\pi_*$ the pull-back and push-forward of $\pi$), and  
$B$ is the magnetic 2-form on the sphere (whose integral over any region gives the magnetic flux through it). Our problem is thus to quantize a phase space with topology $T^*S^2$ and symplectic form given by (\ref{sym}).

This paper is organized as follows. In section \ref{sec:Isham}, we begin with a brief review of Isham's quantization scheme. In section \ref{sec:Quant}, we study the phase space $T^*S^2$ with the charged symplectic form (\ref{sym}), identifying the appropriate quantizing group. In section \ref{ssec:QCO}, we proceed to the quantization, establishing the correspondence between classical observables and quantum self-adjoint operators. In section \ref{ssec:CasInv}, we show that the Casimir invariants of the algebra play an important role in linking the classical and quantum worlds. In particular, this is how the magnetic monopole makes its way into the quantum theory. Next, in section \ref{ssec:Reps}, we study the representations of the group by constructing ladder operators for $J^2$, the angular momentum squared. From the assumption that the theory is free of negative-norm states, in section \ref{ssec:DiracC} we recover Dirac's charge quantization condition. Finally, in section \ref{ssec:Energy}, we compute the energy spectrum for a (non-relativistic) particle on a geometric sphere. In Appendix
\ref{app:Euc} we establish 
the absence of nontrivial central extensions of the Euclidean algebra
${\cal E}_3$, and in Appendix
\ref{app:Alg} we give some details omitted in the derivation of section \ref{ssec:Reps}. In Appendix \ref{app:Bundle} we show how the representation in terms of ``twisted'' wavefunctions can be recovered and, in particular, how the magnetic monopole  is related to their twisting, and
in Appendix \ref{app:ih} we consider non-uniform magnetic fields.

\section{Isham's quantization scheme}
\label{sec:Isham}

In the 
founding days of quantum mechanics,
Dirac 
remarked that ``{\sl The correspondence between the quantum
and classical theories lies not so much the limiting agreement when $\hbar\rightarrow0$
as in the fact that the mathematical operations on the two theories obey in many cases
the same laws.}''~\cite{dirac1925fundamental}. This observation led him to 
postulate the 
general canonical 
quantization scheme which replaces Poisson brackets of classical functions on phase space by quantum commutators of quantum operators, i.e., $[\hat f, \hat g] = i \hbar \widehat{\{ f, g \}}$. More precisely, 
one seeks a linear homomorphism from 
the algebra of 
real functions on the phase space, 
with product defined by the Poisson bracket 
$\{ f, g \}$, to an algebra of self-adjoint operators 
on some Hilbert space, with product defined by the commutator 
$\frac{1}{i\hbar}[\hat f, \hat g]$,
satisfying certain conditions, such as mapping the constant function $f = 1$ to the identity $\hat 1$ and, for all functions $\phi$, mapping $\phi(f)$ to $\phi(\hat f)$. 

It turns out that no such map exists in general, as a consequence of Groenewold--Van Hove obstructions.\footnote{Strictly speaking, the Van Hove no-go theorem applies only for trivial phase spaces, $\ca P = \bb R^{2n}$. The result has been extended to other cases, and it is expected that this kind of obstruction is generic \cite{gotay2000obstructions}. However, some examples have been found where a full, unobstructed quantization is possible \cite{gotay1995full, gotay2001quantizing}.}
Hence one must be careful to select a relatively small set of observables that can be consistently quantized, but which is still large enough to allow for the construction of quantized versions of all other classical observables. The trivial example is that of $\ca P = \bb R^{2n}$, where one can canonically quantize the global coordinates $q^i$ and $p_i$, and then carry along all other observables $f(q, p) \rightarrow f(\hat q, \hat p)$ to the quantum theory (modulo operator-ordering issues). Isham's proposal is to generalize this by identifying 
a transitive group of symplectic symmetries of the phase space
and using it to generate both a special set of classical observables and their associated quantum self-adjoint operators. 
We call this group the {\it quantizing group}. (It is sometimes referred to as the ``canonical group''.)
In the trivial case just mentioned, for example, it could be the group of coordinate translations, 
$(q,p) \rightarrow (q + a, p + b)$, where $q$ and $p$ are any global canonical coordinates. Typically, the dynamical
system possesses other structures that would select a preferred canonical group, such as a metric on the configuration space that appears in the Hamiltonian.
We shall briefly review Isham's scheme in this section. For more details, see \cite{isham1984topological, isham1989canonical}.

Consider a phase space $\ca P$ with symplectic 2-form $\omega$. Assume that the phase space is a homogeneous space for some Lie group $G$ of symplectic symmetries. That is, there is a transitive left action $\delta_g:\ca P \rightarrow \ca P$ of $G$ on $\ca P$ such that $\delta_g^* \omega = \omega$ for all $g \in G$. Each element $\xi$ in the Lie algebra $\al g \sim T_1 G$ 
of $G$ induces a vector field $X_\xi$ on $\ca P$ defined by $X_\xi |_p = \phi_{p *}(\xi)$, where $\phi_p : G \rightarrow \ca P$ is defined by $\phi_p(g) = \delta_g(p)$.
This map is an antihomomorphism from $\al g$ into the algebra of vector fields on $\ca P$, i.e., $[X_\xi, X_\eta] = X_{[\eta, \xi]}$. 
Because $\delta_g$ preserves $\omega$, $X_\xi$ is a (locally) Hamiltonian field, i.e., $\pounds_{X_\xi} \omega = 0$. We therefore have $d(\ii_{X_\xi} \omega) = \pounds_{X_\xi} \omega - \ii_{X_\xi} d\omega = 0$, where $\ii$ denotes the interior product. That is, $\ii_{X_\xi} \omega$ is closed and thus locally exact, so that $dQ_\xi = -\ii_{X_\xi} \omega$ admits local solutions $Q_\xi$, defined up to addition of a constant function on $\ca P$. Since we want these charges\footnote{We call {\sl charge} any function generated in this way by the group of symplectic symmetries, regardless if they are (also) dynamical symmetries in the sense of Noether's theorem.} $Q_\xi$ to play the role of the canonical observables, we require that $G$ generates only {\sl globally} Hamiltonian fields on $\ca P$, meaning that the associated charges
are all defined globally  on $\ca P$.

The symplectic form endows the space of functions on the phase space with an algebraic structure, $\ca A_C$, where the product is given by the Poisson bracket\footnote{Because $\omega$ is non-degenerate, any function $f$ on $\ca P$ can be associated with a unique vector field $X_f$ on $\ca P$ via the relation 
$df = - \ii_{X_f}\omega$. The Poisson bracket between two functions, $f$ and $f'$, is defined by $\{f, f'\} := - \omega(X_f, X_{f'})$.}. In particular, when the functions are taken to be the charges, we have $\{Q_\xi, Q_\eta \} = - \omega(X_\xi, X_\eta)$, and it happens that the map $\xi \mapsto Q_\xi$ is a homomorphism from $\al g$ into $\ca A_C$ up to central charges, that is, $\{Q_\xi, Q_\eta \} = Q_{[\xi, \eta]} + z(\xi, \eta)$, where $z(\xi, \eta)$ is constant on $\ca P$. In practice, we can assume
that this is a true homomorphism, i.e., $z = 0$, since  the group can always be extended 
by a central element to make that so.\footnote{If the central charge $z(\xi, \eta)$ is not trivial (i.e., it cannot be removed by a redefinition of the charges $Q_\xi \rightarrow Q_\xi + f(\xi)$, for some $f: \al g \rightarrow \bb R$), one can always extend the group by a central element so that the extended algebra, $\al g \oplus_S \bb R$, has product law $[(\xi, a), (\eta, b)] = ([\xi, \eta], z(\xi, \eta))$. The new group has a natural action on $\ca P$ (where the central element acts trivially), and the new charges are related to the old ones simply by $Q_{(\xi, a)} = Q_\xi + a$. Consequently, the map $(\xi, a) \mapsto Q_{(\xi, a)}$ is a true homomorphism. Hence, we can always assume that $G$ is already the extension of whatever group we started with.\label{fncentral}}

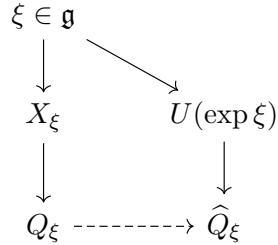
\begin{figure}[h!]
\begin{center}
\begin{tikzcd}
\xi \in \al g \arrow{d} \arrow{rd} & \\
X_\xi  \arrow{d}  & U(\exp \xi) \arrow{d} \\
Q_\xi   \arrow[dashrightarrow]{r} & \widehat Q_\xi
\end{tikzcd}
\end{center}
\caption{\small On the classical side, 
each element $\xi$ of the Lie algebra of
the quantizing group induces a Hamiltonian vector field $X_\xi$ (on the phase space), which in turn defines a Hamiltonian charge $Q_\xi$. On the quantum side, the group element $\exp \xi$ is represented by a unitary transformation (on a Hilbert space), whose self-adjoint generator is $\widehat Q_\xi$.}
\label{diag}
\end{figure}

Next we use $G$ to construct the quantum theory. Let $U: G \rightarrow \text{Aut}(\ca H)$ be  
an irreducible unitary representation of $G$ on 
a Hilbert space $\ca H$. Each element of the algebra $\xi \in \al g$ can be exponentiated to a one-parameter subgroup of $G$, $\exp(t \xi)$, and 
the corresponding one-parameter unitary group is generated
by a self-adjoint operator $\widehat Q_\xi$ on $\ca H$,  as $U(\exp t\xi) = e^{t\widehat Q_\xi/i\hbar}$. From the definition of a representation, the map $\xi \mapsto \widehat Q_\xi$ is a homomorphism from $\al g$ into $\ca A_Q$, where $\ca A_Q$ is the algebra of self-adjoint operators on $\ca H$ with product given by $\frac{1}{i\hbar}[\cdot,\cdot]$.
It follows that the quantization map, which associates to each classical charge $Q_\xi$ the corresponding generator $\widehat Q_\xi$ of the unitary representation,
\begin{equation}
Q_\xi \mapsto \widehat Q_\xi \,,
\end{equation}
is a homomorphism from $\ca A_C$ into $\ca A_Q$.
The logic of this quantization
scheme is summarized in Fig.~\ref{diag}.

Since the space of physical states is actually the ray space, $\ca R := \ca H/U(1)$, corresponding to the quotient of the Hilbert space $\ca H$ by phases $e^{i\theta} \in U(1)$, it is natural to consider also {\it projective} representations of the group $G$. A projective representation is a homomorphism from $G$ into the group of projective unitary operators on $\ca R$, $P\scr U(\ca H)$, consisting of equivalence classes $U \sim e^{i\theta}U$ of unitary operators on $\ca H$. In essence, including projective irreducible unitary representations of $G$ amounts to constructing the quantum theory based on irreducible unitary representations\footnote{Here ``unitary'' means that the observables $Q_\xi$ are represented by self-adjoint operators $\widehat Q_\xi$ on the Hilbert space.}
of the algebra of observables $Q_\xi$.\footnote{Technically, there would also be projective representations of $G$ associated with non-trivial central extensions (by 2-cocycles) of its algebra, if those exist. However, unless such a central extension appears already in the classical Poisson algebra, it will not be of interest to us here, because of the Casimir correspondence principle  introduced below.}

The condition that the group acts transitively on the phase space ensures that any function on $\ca P$ can (locally) be expressed in terms of the canonical observables $Q_\xi$'s. To see this, consider the momentum map, $J$, which is a function from $\ca P$ to $\al g^*$ (the dual algebra of $G$) defined by $J(p)(\xi) := Q_{\xi}(p)$, where $p \in \ca P$ and $\xi \in \al g$. If this map were an {\sl embedding}, then all real functions on $\ca P$ could be written as functions of $Q_\xi$'s. More concretely, note that a basis $\{\xi_i\}$ of $\al g$ induces a coordinate system in $\al g^*$ defined by coordinate functions $w_i(\sigma) := \sigma(\xi_i)$, where $\sigma \in \al g^*$, and these have the property that $Q_{\xi_i} = J^*w_i$. Since any smooth function $f:\ca P \rightarrow \bb R$ could, in this case,  be seen as the pull-back under $J$ of some function $F: \al g^* \rightarrow \bb R$, then $f$ could be written as function of the $Q_{\xi_i}$'s. In general, 
although $J$ may not be an embedding,
transitivity of the group action guarantees that it 
is an {\sl immersion} of $\ca P$ into $\al g^*$.  
Transitivity implies that, at any $p \in \ca P$, 
any tangent vector $V \in T_p \ca P$ is equal to $X_\eta$ for some $\eta \in \al g$. The non-degeneracy of $\omega$ then implies that $dQ_\xi (V) = -\omega(X_\xi, X_\eta)$ is nonvanishing for at least one $\xi$. In other words, there is no direction $V$ along which all charges are (locally) constant, 
implying that any function on $\ca P$ can be locally written in terms of the charges\footnote{To say
there is no direction $V$ along which all charges are (locally) constant
is equivalent to saying that the derivative $J_*$ is injective, so that 
$J$ is an immersion. A further analysis \cite{isham1984topological} reveals that if the immersion fails to be an embedding, 
it is at worst a covering map, i.e., $\ca P$ is a covering space for its image under $J$.
This limits the extent to which functions on $\ca P$ can fail to be globally expressible in terms of the charges.}. 
Therefore, the special set of observables generated in this way is indeed not too ``small''. 

This completes our review of Isham's quantization scheme.
In pursuit of generality, the scheme refers only to minimal structure required to define  
a ``canonical quantization'', which associates to a certain chosen classical 
Poisson algebra of observables a corresponding quantum algebra of observables.
But in order to fully define a physical quantum theory, a particular representation 
of the algebra must be chosen, and the dynamics must be implemented via a 
quantization of the Hamiltonian. This may require additional physical ingredients 
to be introduced in the quantization. In many cases the choice of a representation 
is restricted by what we shall call a {\it Casimir correspondence principle}. 
A classical Casimir invariant is an observable that Poisson commutes 
with the entire Poisson algebra. If that observable admits a quantization 
(i.e., a choice of operator ordering) that commutes with the entire quantum algebra, 
then it is a quantum Casimir invariant. Such a quantum Casimir is a multiple of the 
identity in each irreducible representation of the quantum algebra, so it takes the same 
value in all states of such a representation. For the classical system to arise from a 
classical limit of the quantum system,  the eigenvalue of the  quantum Casimir 
observable should match the corresponding classical value. This Casimir 
correspondence principle  plays an important role in selecting which irreducible 
representation corresponds to the quantization of a given classical system, 
and it sometimes restricts which classical systems can arise from the classical 
limit of a quantum system. In the present paper, for instance, the possible values 
of the magnetic monopole charge will be restricted by this principle to those 
allowed by the famous Dirac condition.

\section{The phase space for a particle on the sphere}
\label{sec:Quant}

In this section we address the classical part of Isham's quantization scheme for the case of a particle on a 2-sphere in the presence of a magnetic monopole. 
The phase space is $\ca P = T^*S^2$ but, as explained in the introduction, the symplectic form must be given by (\ref{sym}) if the Hamiltonian is to be a globally well-defined function. Our goal is to identify a suitable group of symplectic symmetries of this phase space and then compute the associated Poisson charges. 

\subsection{A transitive group of symplectic symmetries}
\label{ssec:Class}

A magnetic field 2-form $B$ on $S^2$ admits an infinite dimensional symmetry group that acts transitively on $S^2$,
provided that $B$ is nowhere vanishing. This is the group of  
``area'' preserving diffeomorphisms, where $B$ defines the area element. 
We are interested in quantizing an $SO(3)$ subgroup of this group, {which not only is the smallest group that can act transitively on $S^2$ but 
is also the group of isometries of a round metric}. There are infinitely many such 
subgroups, which all lead to equivalent phase space quantizations. We write the magnetic field as
\begin{equation}\label{B}
B = g \epsilon \,,
\end{equation}
where the 2-form $\epsilon$ is scaled so that $\int_{S^2}\epsilon = 4\pi$, and $g$ is a dimensionful coefficient.
The total flux through the sphere is simply $4\pi g$, so $g$ can be interpreted as the magnetic charge of a monopole ``inside''\footnote{If the magnetic field were not nowhere vanishing, we could still separate it as $B = g \epsilon + dA$,
where $A$ is a globally defined potential 1-form that could be included in the Hamiltonian, while the 
$g \epsilon$ term could be included in the symplectic form.}.
The kinetic energy term in the Hamiltonian for a charged particle on $S^2$ 
involves a particular metric on the $S^2$. In order to naturally quantize this Hamiltonian, we will ultimately base the quantization on the $SO(3)$ group of isometries of this metric, however that choice plays no role until we come to quantizing the Hamiltonian.

We denote by $l_R(x)$ the action of a rotation
$R \in SO(3)$ on a point $x$ on the sphere. 
As with any action on the configuration space, there is a natural lifted action to the cotangent bundle, defined by
\begin{equation}\label{liftL}
L_R (p) = l_R^{-1*}p \,,
\end{equation}
which maps the fiber over $x$ to that over $l_R(x)$, i.e.\ it satisfies $\pi \circ L_R = l_{R} \circ \pi$.
The canonical potential 1-form $\theta$ is invariant under the lift of 
{\it any} point transformation (diffeomorphism of the configuration 
space).\footnote{Let $\phi$ be a diffeomorphism of the configuration space, and let $\Phi$ be its lift to the phase space, i.e., $\Phi(p) := \phi^{-1*}p$. If $V \in T_p \ca P$, then $\Phi^*\theta(V) = \theta(\Phi_*V) = (\Phi (p))(\pi_*\Phi_*V) = (\phi^{-1*}p)(\pi_*\Phi_*V) = p(\phi_*^{-1}\pi_*\Phi_*V) = p(\pi_*V) = \theta(V)$.
Alternatively, using coordinates, $p'_i dx'^i = p_j\frac{\partial x^j}{\partial x'^i} \frac{\partial x'^i}{\partial x^k}dx^k = p_k dx^k$.}
In particular, we have $L_R^*\theta=\theta$, and therefore $L_R^* d\theta = dL_R^*\theta = d\theta$. 
Moreover $\pi^*B$ is invariant under rotations:
$L_R^* (\pi^*B) = (\pi \circ L_R)^* B = (l_{R} \circ \pi)^*B = \pi^*B$. 
Therefore the symplectic form \eqref{sym} is invariant,
\begin{equation}\label{omegainv}
L_R^*\omega 
= \omega \,.
\end{equation}
That is, for all $R$, $L_R$ is a symmetry of the symplectic form.

The quantizing group should be larger than just $SO(3)$, since the rotations act only ``horizontally'' on the phase space.  For the quantizing group to act transitively, it should include elements that move points along the fibers of the cotangent bundle. The simplest ``vertical'' action is a translation of momentum, 
\begin{equation}
F_\alpha (p) = p - \alpha \, ,
\end{equation}
where $\alpha$ is a 1-form field on $S^2$.
(For notational simplicity we leave implicit the point $\pi(p)$ at which 
$\a$ is evaluated.)
This acts on the symplectic form \eqref{sym} as
\begin{equation}
F^*_\alpha \omega = d (F^*_\alpha \theta) + e (\pi \circ F_\alpha)^* B  \,.
\end{equation}
The term $\pi^*B$ is invariant since $\pi \circ F_\alpha = \pi$. The symplectic potential, however, transforms non-trivially. For any $V \in T_p \ca P$, we have
\begin{align}
F^*_\alpha \theta (V) = \theta (F_{\alpha *} V) = (F_\alpha p) (\pi_* F_{\alpha *} V) \nonumber\\
= (p - \alpha)(\pi_* V) = (\theta - \pi^*\alpha)(V) \,, \label{Fatheta}
\end{align}
so that 
\beq
F^*_\alpha \omega = \omega - \pi^* d\alpha \,.
\eeq
In order for $F_\alpha$ to be a symplectic symmetry, we must  require $\alpha$ to be closed. We will restrict further to {\sl exact} 1-forms, $\alpha = df$, with $f$ globally defined on $S^2$ 
 to ensure that the associated charges will be globally defined. 

The (infinite-dimensional) space of all exact 1-form fields
is unnecessarily large, so we look for a ``minimal'' set of momentum translations that act transitively along the fibers and are consistent with the spherical symmetry, in the sense that they extend the chosen $SO(3)$ 
into a larger group.
As observed by Isham (in a more general setting), a suitable set can be generated by realizing
the configuration space $S^2$ as
an orbit of a representation of $SO(3)$ in a vector space, and
defining the momentum translations as
the pullback to the orbit of the ``constant''
1-forms on that vector space.
In particular, we can choose the fundamental representation on $\bb R^3$, and
identify the $S^2$ with the orbit passing 
through $u = (0,0,1)$ in $\bb R^3$, that is, with the set of unit vectors 
$x \in \bb R^3$ such that $x = Ru$ for some $R \in SO(3)$.\footnote{This identification of  
the configuration space with the unit sphere in the abstract $\bb R^3$ should not be confused with the physical sphere, which may have its own geometry. The orbit is a ``unit sphere'' with respect to the inner product $\langle v,u \rangle = \sum_{i=1}^3 v_i u_i$ on the abstract $\bb R^3$.}  
(The notation ``$x$'' for
these vectors coincides with that which we used already to label the points in $S^2$.) Any dual vector $\alpha \in \bb R^{3*}$, 
can naturally be seen as a 1-form field on $\bb R^3$. Moreover, this 1-form field is exact, 
for it can be written as $df_\alpha$, where the function $f_\alpha : \bb R^3 \rightarrow \bb R$ is defined by
\begin{equation}\label{falpha}
f_\alpha(x) := \alpha(x)  \,.
\end{equation}
These 1-form fields can be pulled-back to $S^2$ to define the corresponding action along the fibers of $\ca P$. 

Combining the $L_R$ and $F_\alpha$ transformations, we get a transitive group of symplectic symmetries of the phase space:
the semidirect product $G = \bb R^{3*} \rtimes SO(3)$, acting on $\ca P$ as
\begin{equation}\label{Lambda}
\Lambda_{(\alpha, R)} (p) = l_R^{-1*}p - \alpha  \,,
\end{equation}
which satisfies the product rule
\begin{equation}\label{Glaw}
(\alpha, R)(\alpha', R') = (\alpha + l_R^{-1*}\alpha', RR')  \,.
\end{equation}
Since $\bb R^{3*} \sim \bb R^{3}$ and the co-representation of $SO(3)$ in $\bb R^{3*}$ is equivalent to its representation in $\bb R^{3}$,\footnote{In a matrix realization, $R^{-1*}\alpha = (R^{-1})^T\alpha = R\alpha$.} this group is isomorphic to the Euclidean group, $E_3 = \bb R^{3} \rtimes SO(3)$.

We take this group, $G$, to be the quantizing group. Note that it is independent of the magnetic term in the symplectic form. Since its algebra does not admit any non-trivial central extension by 2-cocycles $z(\xi, \eta)$,\footnote{Since we found no explicit demonstration of this statement in the literature, we include one in Appendix \ref{app:Euc} for completeness. This formal proof is not really necessary for our purposes, however, as in the next section we explicitly compute the Poisson algebra and show that no central charges arise.}
no (non-trivial) central charge can appear in the associated Poisson algebra. The {\sl quantizing algebra} is thus the same as in the uncharged case. However, 
as we shall see in Sec.~\ref{ssec:CasInv},
the magnetic term makes itself felt through the value of a Casimir invariant of the corresponding Poisson bracket algebra, which carries over to the quantum theory.

\subsection{Classical canonical observables}
\label{ssec:CCO}

We next compute the classical charges associated with the quantizing group $G$,
beginning with the $SO(3)$ generators.  Let $n$ be an element of the algebra $\al{so}(3)$, and denote its exponential by $R_n = \exp(n)$.
(Here $\exp:\al g\rightarrow G$ is the usual Lie group exponential map.)
{Let $\overline{X}_n$ be the vector field (on $S^2$) induced by $n$ through the action of $SO(3)$ on $S^2$, and let $X_n$ be the vector field (on $\ca P$) induced by $n$ through the lifted action of $SO(3)$ on $\ca P$ defined in \eqref{liftL}. Since the group action on $\ca P = T^*S^2$ maps fibers into fibers, we note that $\overline{X}_n$ is just the projection of $X_n$ to the sphere, i.e.,  $\overline{X}_n = \pi_* X_n$.}
The corresponding charge $P_n$ is defined by
\begin{align}
dP_n &= - \ii_{ X_n}\omega \nonumber\\
& =- \ii_{ X_n}(d\theta + eg\,\pi^*\epsilon) \nonumber\\
&=d[\theta(X_n)] - eg\,\pi^*\ii_{\overline{X}_n}\epsilon  \,. \label{3rd}
\end{align}
The first term in the last line follows from $0 = \pounds_{ X_n} \theta = \ii_{ X_n}d\theta + d \ii_{ X_n}\theta$, and 
$\theta$ is invariant under any point transformation, as discussed in the paragraph leading to (\ref{omegainv}). Like the first term, the second term is also an exact 1-form:  it is closed since
$d(\pi^*\ii_{\overline{X}_n}\epsilon)
=\pi^*d\ii_{\overline{X}_n}\epsilon
=\pi^*{\cal L}_{\overline{X}_n}\epsilon=0$, and since $S^2$ is simply connected it is therefore also exact, i.e., $\ii_{\overline{X}_n}\epsilon=d\Gamma_n$, for some $\Gamma_n : S^2 \rightarrow \bb R$.
The charge $P_n$ is thus given by
\begin{equation}
P_n =p(\overline{X}_n) - eg\,\Gamma_n\circ \pi 
\end{equation}
up to an additive constant. The use of the symbol ``$P$'' is motivated by the fact that the charges associated with spatial transformations are the analogue of momentum coordinates. The term $p(\overline{X}_{n})$ alone is 
the usual orbital angular momentum associated with the rotation Killing vector field $\overline{X}_{n}$,
while $P_n$ is the canonical angular momentum, i.e., the charge that generates rotations on the phase space
with symplectic form $\omega$ \eqref{sym}. 
On a phase space with the ``usual'' symplectic form $d\theta$, 
the canonical angular momentum would have been simply $p(\overline{X}_{n})$.

Next we compute the charges associated with the $\bb R^{3*}$ part of the group. Since the group $\bb R^{3*}$ is a vector space, it can be naturally identified with its Lie algebra.
Let $Y_\alpha$ be the momentum 
translation vector field on $\ca P$ induced by an element $\alpha$ of the Lie algebra
of $\bb R^{3*}$ .
The corresponding charge $Q_\a$ is defined by 
\begin{align}\label{Qaderivation}
dQ_\a &= - \ii_{Y_\a}\omega \nonumber\\
&=- \ii_{Y_\a}d\theta \nonumber\\
&= - \pounds_{Y_\alpha}\theta \nonumber\\
&= - \left. \frac{d}{dt} F_{t\alpha}^* \theta \right|_{t=0} \nonumber\\
&= \pi^* \alpha \nonumber\\
&= \pi^* df_\alpha \nonumber\\
&= d(f_\a\circ\pi) \,,
\end{align}
where in the second line we used that $\ii_{Y_\a} \epsilon = 0$; in the third line that $\ii_{Y_\alpha}\theta = 0$; in the fourth line that the flow induced by $\alpha$ is $p \mapsto F_{t\alpha}(p)$; in the fifth line we used (\ref{Fatheta}); and in the sixth line we used that $df_\alpha = \alpha$, as defined in \eqref{falpha}.\footnote{Alternatively, in local coordinates adapted to the bundle structure of $T^*S^2$, the flow generated by $\a = \a_i dq^i$ is given by $F_{t\a}(q^i, p_i) = (q^i, p_i - t\a_i)$, so $Y_\a = - \a_i \frac{\partial}{\partial p_i}$, and the charge differential is  $dQ_\a = - i_{Y_\a} (dp_i \wedge dq^i + eg\, \epsilon_{ij} dq^i \wedge dq^j) = \a_i dq^i = \pi^*\a$. In the last step the $\pi^*$ appears because, in this equation, $q^i$ are coordinates on $T^*S^2$, while in the definition of $\alpha$ they are coordinates on $S^2$.}
The charge associated with $\a$ is therefore
\begin{equation}\label{Qa0}
Q_\alpha =   f_\alpha \circ \pi \,
\end{equation}
up to an additive constant.
The use of the symbol ``$Q$'' here is motivated by the fact that the charges associated with momentum
translations are the analogue of position coordinates. 

To be more concrete, it is convenient to use the realization of $S^2$ as the unit sphere in the abstract $\bb R^3$, 
which was introduced in the previous subsection.
We identify $n \in \al{so}(3)$ with the vector in $\bb R^3$ whose direction 
is the corresponding axis of rotation and whose magnitude $|n|$ gives 
the angle of rotation of $\exp(n)$, according to the right-hand rule.
Then, using adapted spherical coordinates
in which $n$ is aligned with $\theta=0$, we have
$\overline{X}_n=|n|\partial_\phi$, hence  
$\ii_{\overline{X}_n}\epsilon=|n| \, \ii_{\partial_\phi}\! \sin\theta\, d\theta\wedge d\phi=
|n|\,d(\cos\theta)= d(n\cdot x)$, so we can choose $\Gamma_n = n \cdot x$. Therefore, the canonical charges are given, up to an additive constant, by
\begin{align}
&P_n = p(\overline{X}_n) - eg\, n \cdot x \label{Pn1}\\
&Q_\alpha = \a(x)  \,. \label{Qa1}
\end{align}
The notation is somewhat abbreviated here. 
Strictly speaking, $P_n$ and $Q_\a$ are functions on the phase space ${\cal P}$, which is
specified above by giving their values at a point $p\in {\cal P}$. The vector field 
$\overline{X}_n$ is implicitly evaluated at $\pi(p)$, and $x$ is the unit vector representative
of $\pi(p)$ in the embedded realization of $S^2\subset \bb R^3$.

We next consider the Poisson bracket algebra of the charges. 
By construction, this algebra matches the 
Lie algebra of the canonical group that defined the charges, up to a possible central extension.
If a central extension appears in such an algebra, in general
it may or may not be removable using the freedom
to shift the charges by addition of constants. 
As mentioned above, the Euclidean algebra in itself (i.e.\ apart from 
any canonical realization)
does not admit any non-trivial central extension (by 2-cocycles), so that it must be 
possible to choose the additive constants such that the Poisson algebra matches 
the Lie algebra. In fact, the choices we have made in \eqref{Pn1} and \eqref{Qa1}
satisfy this criterion, and the Poisson algebra 
takes the form
\begin{align}
&\{P_n,  P_{n'}\} =P_{[n,n']} \nonumber\\
&\{ Q_\a, P_n\} =Q_{{\cal L}_{X_n}\a} \nonumber\\
&\{Q_\a,Q_{\a'}\} = 0 \label{algJN0} \,,
\end{align}
 which matches the semi-direct product structure of the algebra 
$\al g = \bb R^{3*} \oplus_S \al{so}(3)$ of $G$
without central charges. That is, denoting elements of $\al g$ by $(\alpha, n) \in \bb R^{3*} \oplus_S \al{so}(3)$, the product rule reads
\begin{align}
&[(0, n), (0, n')] = (0, [n, n']) \nonumber\\
&[(\alpha, 0), (0, n)] = (\ca L_{X_n}\alpha, 0) \nonumber\\
&[(\alpha, 0), (\alpha', 0)] = 0  \,,
\end{align}
revealing how the linear association $(\a, n) \mapsto P_n + Q_\a$ is a (true) homomorphism. 

To verify that the choices \eqref{Pn1} and \eqref{Qa1} lead to no central charges, and for later purposes, 
it is convenient to introduce a basis for $\al g$. Using the identification $\al{so}(3) \sim \bb R^3$, choose an orthonormal basis $\{e_i\}$ ($i = 1,2,3$) in $\bb R^3$. (Note that $\exp(e_i)$ implements a right-handed rotation by the angle $1$ around the $i$-axis.) It is straightforward to check that $[e_i,e_j]=\varepsilon_{ijk} e_k$, 
where $\varepsilon_{ijk}$ is the Levi-Civita symbol.\footnote{\label{fn12} As $SO(3)$ acts on $\bb R^3$ from the left, the algebra element $n \in \bb R^3$ induces the vector field $\left. \overline{X}_n \right|_x = n \times x$, where $x \in \bb R^3$. The Lie bracket of two such vector fields is given by $[\overline{X}_n, \overline{X}_{n'}] = - \overline{X}_{n \times n'}$. Together with  $[\overline{X}_\xi, \overline{X}_\eta] = \overline{X}_{[\eta, \xi]}$ (see third paragraph of section \ref{sec:Isham}) this yields $[n, n'] = n \times n'$.}
Let $\{ e^i \}$ denote the dual basis, satisfying $e^i(e_j) = \delta^i_{\,\, j}$. We define
\begin{align}\label{JiNidef}
&J_i := P_{e_i} \nonumber\\
&N_i := Q_{e^i} \,,
\end{align}
which satisfy the algebra
\begin{align}
&\{J_i,  J_j\} =\varepsilon_{ijk}  J_k \nonumber\\
&\{ J_i,  N_j\} = \varepsilon_{ijk}  N_k \nonumber\\
&\{ N_i,  N_j\} = 0 \label{algJN}\,.
\end{align}
This  is the algebra of the Euclidean group, presented in terms of a basis of generators of rotation and translation.

We can express (\ref{Pn1}) and (\ref{Qa1}) in this basis. 
If $x \in S^2 \subset \bb R^3$, we can write $\overline{X}_{e_i} =e_i \times x$. Also, if $p$ is a co-vector on $S^2$ at $x$, we can (abusing the notation) associate it with a vector $p$ in $\bb R^3$, tangent to $S^2$ at $x$, such that $p \cdot v = p(v)$, where $v$ is any vector on $\bb R^3$ tangent to $S^2$ at $x$. In this way, we have $p(\overline{X}_{e_i}) = p \cdot (e_i \times x) = e_i \cdot (x \times p)$, 
which is the familiar orbital angular momentum about the 
axis $e_i$ in $\bb R^3$. Thus, $J_i = e_i \cdot ( x \times p - eg\, x )$.
Also, $N_i$, evaluated at any point in the fiber over $x$, can be written as $N_i = e^i(x) = e_i \cdot x$. In a 3-vector notation,
\begin{align}
J &= x \times p - eg\, x \label{Jivec}\\
N &= x \,,\label{Nivec}
\end{align}
so we have $J_i = e_i \cdot J$ and $N_i = e_i \cdot N$. 

To establish \eqref{algJN}, i.e., to verify 
that indeed there are no missing central terms, we may evaluate the
brackets at points in the phase space where both sides of 
the equation vanish. For example, 
recall that $\{J_1,J_2\}= -\omega( X_{e_1}, X_{e_2})$. 
The vector field $ X_{e_1}$ vanishes at the points in phase space
with zero momentum and located at the rotation axis
$e_1$ on the $S^2$ (i.e., the two points in the intersection of the zero section with the fibers over $x = \pm e_1$).
Hence $\{J_1,J_2\}$ vanishes there.
On the other hand, according to \eqref{Jivec}, at the same point
the function $J_3$ is equal to $-eg\, e_3\cdot e_1=0$. 
Any constant added to $J_3$ would spoil this agreement. This argument 
works for all of the $J_i$ brackets, so we conclude that 
no central term need be added on the right hand side of the first bracket in
\eqref{algJN}. The argument just given 
also implies that $\{J_1,N_2\} = -\omega( X_{e_1}, Y_{e^2})$ vanishes at the axis point $e_1$, while 
$N_3$ equals $e^3( e_1)=0$ at that same point. Hence no central 
term appears in the second bracket either. As for the last brackets 
in \eqref{algJN}, since the right hand side vanishes, it is unaffected by 
addition of a constant to any charge. In conclusion, as we claimed, the charges defined in \eqref{Pn1} and \eqref{Qa1} provide a realization of the quantizing algebra without central charges.

\section{The quantum theory}
\label{sec:Quantum}

{Quantization of the theory amounts to constructing 
a unitary irreducible (UI) projective representation of the canonical group $G \sim E_3$,
in which the value 
of all classical Casimir invariants carry over to the quantum theory, 
modulo possible operator ordering ambiguity 
that might arise in quantizing the Casimir invariant.
Since the Euclidean algebra does not admit (non-trivial) central extensions, the UI 
projective representations of $E_3$ are in correspondence with true
UI representations of its universal cover, $\widehat E_3 \sim \bb R^3 \rtimes SU(2)$,
which in turn are in correspondence with UI representations of the Euclidean algebra $\al g = \bb R^3 \oplus_S \al{so}(3)$.}\footnote{See \cite{bargmann1954unitary}; or, for an informal discussion, \cite{StackEx}.}

\subsection{Quantum canonical observables}
\label{ssec:QCO}

The quantum version of the classical canonical observables $J_i$ and $N_i$ are the self-adjoint generators of the corresponding unitary transformations in some Hilbert space $\ca H$. That is, given some unitary irreducible representation $U$ of $G$ (or its universal cover), and denoting elements of $\al g$ by $(\alpha, n)$, we define the operators $\widehat J_i$ and $\widehat N_i$ by
\begin{align}
&U[\exp(0,\lambda e_i)] =: e^{-i \lambda \widehat J_i/\hbar} \nonumber\\
&U[\exp(\lambda e^i,0)] =: e^{-i \lambda \widehat N_i/\hbar}  \,, \label{qexp}
\end{align}
where $\lambda$ is an arbitrary real parameter.
It follows from the group structure that the quantized algebra satisfies
\begin{align}
&[\widehat J_i, \widehat J_j] = i\hbar\, \varepsilon_{ijk} \widehat J_k \nonumber\\
&[\widehat J_i, \widehat N_j] = i\hbar\, \varepsilon_{ijk} \widehat N_k \nonumber\\
&[\widehat N_i, \widehat N_j] = 0 \,. \label{EucAlg}
\end{align}
The quantization map is 
$J_i \rightarrow \widehat J_i$ and $N_i \rightarrow \widehat N_i$,
and \eqref{EucAlg} is the quantization of the Poisson algebra \eqref{algJN}. 
 For notational simplicity we shall henceforth omit the ``hat'' symbol over quantum operators, 
 since it should be clear from the context whether we are referring to the classical or the quantum observables. 

\subsection{Casimir invariants}
\label{ssec:CasInv}

Casimir operators, by definition, commute with all elements of the algebra, and their eigenvalues
can therefore be used to label its irreducible representations\footnote{If an operator commutes with all other operators in an irreducible representation, then according to Schur's lemma it must be proportional to the identity operator.}. There are two independent Casimir operators associated with the algebra $\al g$,
\begin{align}
&N^2 = \sum_i (N_i)^2 \nonumber\\
&N \cdot J = \sum_i N_i J_i  \,. \label{QCasimir}
\end{align}
Their classical correspondents Poisson-commute with 
everything in the classical algebra and, using \eqref{Nivec} and \eqref{Jivec},
we have 
\begin{align}
&N^2 = x^2 = 1 \nonumber\\
&N \cdot J = -eg\, x^2 = -eg  \,, \label{CCasimir}
\end{align}
revealing that these two quantities  
are constant classical observables. 
Since no operator ordering ambiguities arise in quantizing $N^2$ and $N\cdot J$ ($= J \cdot N$),  we take as part of the quantization prescription that the quantum theory must carry the representation for which the values of the two Casimirs (\ref{QCasimir}) are given by precisely the corresponding classical values (\ref{CCasimir}). 
Note that the value of $N\cdot J$ is the {\it only} 
way the presence of the magnetic monopole is felt in the quantum theory.

\subsection{Representations of the algebra}
\label{ssec:Reps}

{While the representation theory of the Euclidean group is well known from Mackey's theory of induced representations carried by a space of wavefunctions~\cite{niederer1974realizations, raczka1986theory}, 
we will present it here using a rather simpler abstract  ladder-operator approach.}\footnote{After completing this
work we found that a similar realization of the representation appears in \cite{kowalski2000quantum}, although no explicit derivation is presented there.} 
The method is reminiscent of the usual derivation of the unitary irreducible representations of $SU(2)$. 
Some of the details are left to Appendix \ref{app:Alg}. In Appendix \ref{app:Bundle} we discuss an alternative derivation based on Mackey theory, which provides a construction of the Hilbert space based on wave functions on $S^2$.  

Note first that the Euclidean algebra (\ref{EucAlg}) is invariant under rescaling of $N$. Thus, without loss of generality, we can particularize to the representation with $N^2=1$. The value of the other Casimir is left arbitrary,
\begin{equation}\label{NJs}
N \cdot J = s\hbar \,,
\end{equation}
with $s$ some real parameter.

Let us start with a basis of simultaneous eigenvectors of $J^2$ and $J_3$, denoted as $|j,m\rangle$, defined by
\begin{align}
&J^2|j,m\rangle = j(j+1) \hbar^2 |j,m\rangle \nonumber\\
&J_3|j,m\rangle = m  \hbar|j,m\rangle  \,.
\end{align}
At this point it is not clear which values of $j$ are allowed, for a given $s$,  
nor if there is more than one state with a given value of $j$ and $m$.
Note that $J_\pm = J_1 \pm i J_2$ act as raising and lowering operators for $m$ and, from the standard analysis of the angular momentum algebra, we know that, for a given $j$, $m$ varies from $-j$ to $j$ in integer steps. Also, $j$ can only be a non-negative integer or half-integer {(i.e., $j \in \frac{1}{2}\bb Z^{0+}$)}, but it may be that only a subset of that is included. 

Before systematically deriving the properties of the irreducible representations, 
it is enlightening to guess, by a simple but non-rigorous 
reasoning, which values of $s$ and $j$ are included. Supposing that there is (in a limiting sense)  a state  localized at the north pole of the sphere, 
the operator $N\cdot J$ acts on such a state as $J_3$ (see \eqref{u-state} for details). This implies that $2s$ must be integer valued in order for the representation to be non-trivial. Since it has the $J_3$ eigenvalue $s\hbar$, 
such a state must be constructed from states
with $j\ge s$. By virtue of rotational symmetry, the same can be said about  states localized at any other point on the $S^2$, and one would expect an irreducible representation to be constructed from the span of these states
with $j\ge s$.
Moreover, since $N$ is a vector operator 
(in the way it transforms under commutation with the $J_i$), 
its action can change the value of $j$ by plus or minus unity,
which suggests that repeated action of $N$ will both raise the $j$ values without bound and lower them 
until they reach a floor, which presumably lies at $j=s$, since $j\ge s$ is the only apparent constraint.
That is, the representation 
must include all values $j=s+n$, 
for non-negative integers $n$.
Indeed this is the correct 
spectrum, as we now
show by explicit construction of the representation.\footnote{The earliest mention of this spectrum that
we have found appears in  \cite{haldane1983fractional},
although no derivation was given there.}

We now present the rigorous derivation of
the irreducible representations, 
analyzing how certain operators in the algebra act as ``ladder operators'' to 
shift the values $j$ and $m$.
From the Wigner-Eckart theorem we know that when 
the vector operator $N_i$ acts
on a state,
it can only change 
the value of $j$ by $-1$, $0$ or $1$. 
And, since acting on a state with 
$N_\pm = N_1 \pm i N_2$ changes the $m$ value by $\pm1$, it follows that
$N_+ |j,j\rangle \propto |j+1,j+1\rangle$. Thus $N_+$ acts as a raising operator for {\sl edge} states $|j,j\rangle$, i.e.\ states with the maximal $m=j$ for a given $j$. For now, let us assume that if $|j,j\rangle$ is in the Hilbert space then so is $N_+|j,j\rangle$, i.e., that its norm is positive.  This assumption will be justified later. 

Now let $|j_0, j_0\rangle$ be the {\sl ground} state, in the sense that there are no states with $j < j_0$ in the representation being constructed. (We know that there must be such a lowest $j$ state, since $j$ is non-negative.) Since we are constructing an irreducible representation, the whole Hilbert space $\ca H$ must be generated by acting with all elements of the algebra on any given state, in particular the ground state. Consider then the set of states
\begin{equation}\label{basis}
|j,m\rangle := (J_-)^{j-m} (N_+)^{j-j_0}|j_0,j_0\rangle  \,,
\end{equation}
where $j - j_0 \in \bb Z^{0+}$ and $-j \le m \le j$. 
Since these states have distinct eigenvalues for the the self-adjoint operators $J^2$ and/or $J_3$, they 
are necessarily orthogonal (although not normalized). 
We will show that a representation exists only if
$s \in \frac{1}{2}\bb Z$, in which case 
there is a unique irreducible representation, spanned by the states \eqref{basis}
with $j_0= |s|$.
To establish this, we first prove that these states 
are closed under the action of the entire algebra, and then we show that that they all have positive norm,
provided the $s$ quantization condition holds and $j_0$ has the required value.

\begin{figure}
\centering
\begin{minipage}{.5\textwidth}
  \centering
  \includegraphics[scale = 0.4]{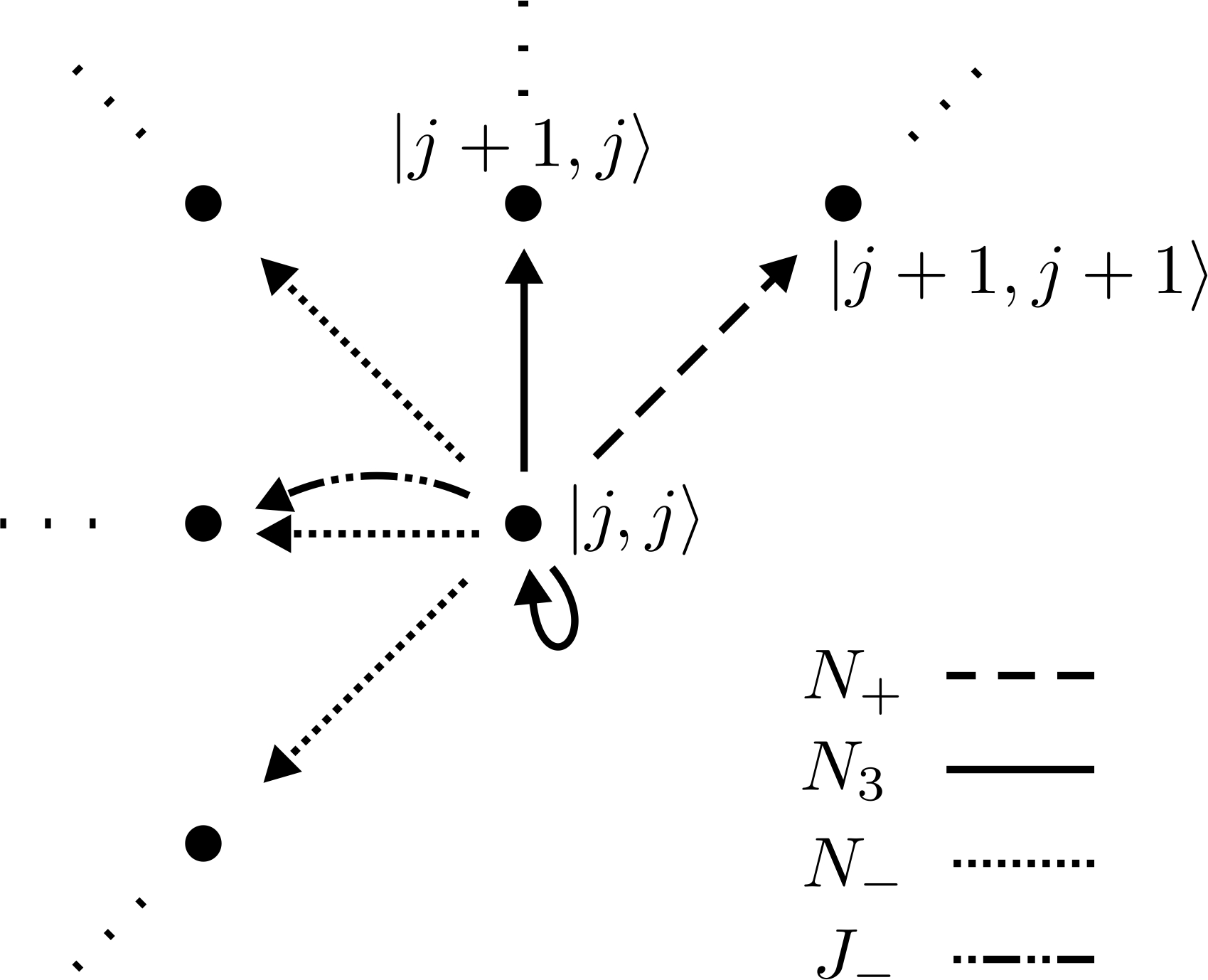}
\end{minipage}%
\begin{minipage}{.5\textwidth}
  \centering
  \includegraphics[scale = 0.4]{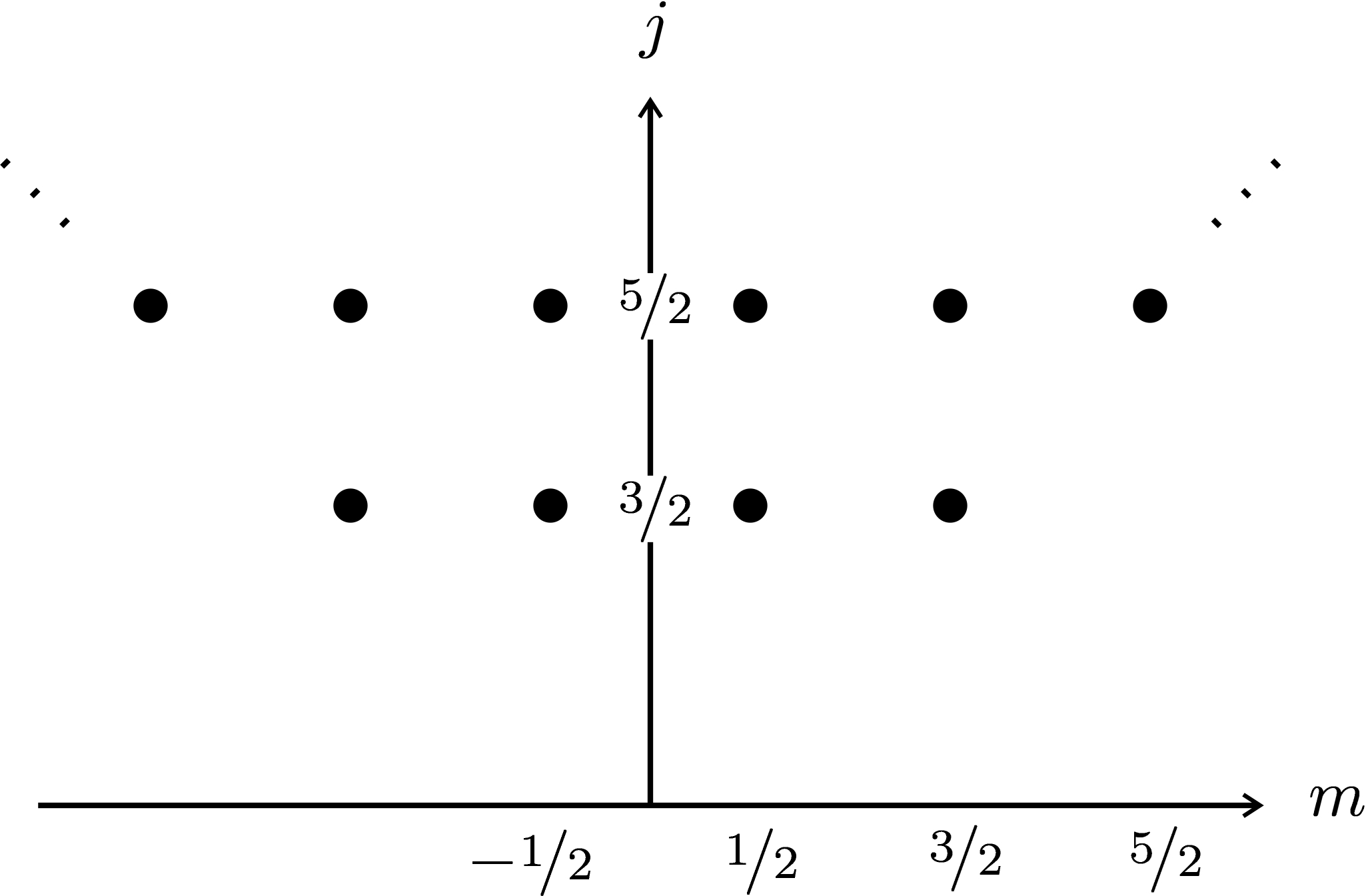}
\end{minipage}
\caption{The diagram on the left depicts an arbitrary edge state $|j, j\ra$ and the action of the operators $N_+$, $N_3$, $N_-$ and $J_-$ on it --- the image of $|j,j\ra$ is in a linear superposition of states at the endpoints of the corresponding arrows; note that $J_3$ keeps $|j, j\ra$ fixed and $J_+$ annihilates it. The diagram on the right shows the example of the representation with $s = 3/2$, indicating the states that are present in it.}
\label{fig:rep}
\end{figure}

It is clear how to define the action of $J_i$ on this basis, using just the angular momentum algebra. That is, $J_i |j,m\rangle$ can be written as a linear combination of $|j, m-1\rangle$, $|j, m\rangle$ and $|j, m+1\rangle$ with coefficients determined by the algebra. So let us focus on defining the action of the $N$'s. It is easy to see that, since the commutator of an $N$ with a $J$ gives an $N$, the action of $N_i$ on any state is well-defined provided that the $N$'s have a well-defined action on the edge states $|j,j\rangle$. Similarly, if the edge states have positive norm then so do the rest of the states.

By the definition of the basis states \eqref{basis} we have
\begin{equation}
N_+ |j,j\rangle= |j+1,j+1\rangle  \,.
\end{equation}
Next, using only the algebra and the Casimir invariants, we show in Appendix \ref{app:Alg} that
\begin{equation}\label{N3}
N_3|j,j\rangle =  \frac{s}{(j+1)} |j,j\rangle  -\frac{1}{2(j+1)} |j+1,j\rangle
\end{equation}
and, for $j > j_0$,
\begin{align}
N_- |j,j\rangle = &\frac{2j}{2j+1} \left( 1 - \frac{s^2}{ j^2} \right) |j-1,j-1\rangle +  \nonumber\\
&~~~~~+s\frac{|j,j-1\rangle} {j(j+1)} -\frac{|j+1,j-1\rangle}{2(2j+1)(j+1)}   \,. \label{N-j}
\end{align}
(The action of the algebra on the edge state $|j, j\ra$ is depicted on the left in Fig.~\ref{fig:rep}.)  It then remains to determine $N_-|j_0, j_0\rangle$,  to 
establish that the states $|j,j\rangle$ with $j\ge j_0$ have positive norm, and to show that 
the algebra is consistent with the assumption that $j$ cannot be lowered below $j_0$. 

Observe that 
the form of (\ref{N-j}) indicates that the operator
\begin{equation}\label{Lj}
L^{(j)} = N_- - \frac{s}{j(j+1)} J_- + \frac{1}{2(2j+1)(j+1)} (J_-)^2 N_+
\end{equation}
acts as a $j$-lowering operator for $|j,j\rangle$, i.e.\ , $L^{(j)}|j,j\ra\propto |j-1,j-1\ra$,
modulo the assumption that $j>j_0$. In fact, 
using only the algebra it can be established for any $j$, including $j_0$, that
\beq
[J^2, L^{(j)}] |j,j\rangle =  -2j\hbar^2  L^{(j)} |j,j\rangle \,,
\eeq
hence  $J^2 L^{(j)} |j,j\rangle = j(j-1)\hbar^2 L^{(j)} |j,j\rangle$.

The squared norms of the raised and lowered edge states 
can be computed using only the algebra and Casimir invariants.
Doing so, we find 
\begin{equation}\label{N-N+}
\|N_+|j,j\ra\|^2 = \frac{2(j+1)}{2j+3} \left( 1 - \frac{s^2}{ (j+1)^2} \right) \la j,j|j,j\ra
\end{equation}
\begin{equation}\label{LdL}
\|L^{(j)}|j,j\ra\|^2 = \frac{2j}{2j+1} \left( 1 - \frac{s^2}{ j^2} \right) \la j,j|j,j\ra \,.
\end{equation}
(See Appendix \ref{app:Alg} for more details.)
Starting with a state $|j,j\ra$, we can apply a sequence of $L^{(j)}$ operators to lower $j$ successively. 
This process stops, preventing $j$ from becoming negative, only if $j_0$ differs from $j$ by an integer and $L^{(j_0)}|j_0, j_0 \rangle = 0$. It thus follows from (\ref{LdL}), and the non-degeneracy of the Hilbert space, that $j_0 = |s|$. Moreover, the condition $L^{(j_0)}|j_0, j_0 \rangle = 0$, together with (\ref{Lj}), defines the action of $N_-$ on the ground state,
\begin{equation}\label{N-j0}
N_- |j_0,j_0\rangle = {s} \frac{|j_0,j_0-1\rangle}{j_0(j_0+1)}-\frac{|j_0+1,j_0-1\rangle}{2(2j_0+1)(j_0+1)}   \,.
\end{equation}
This happens to agree with (\ref{N-j}) particularized to $j = j_0 = |s|$ {(i.e., although the derivation of  (\ref{N-j}) applied only for $j>j_0$, the result actually extends to $j=j_0$).}
Finally,  (\ref{N-N+}) shows that, since $j\ge |s|$,  $N_+$ always creates a positive-norm state.
This establishes that, if $|j_0,j_0\rangle$ has positive (squared) norm, then so do all 
of the other edge states.
It follows that all states defined in (\ref{basis}) have positive norm and thus must be in $\ca H$. (This array of states is illustrated
on the right in Fig.~\ref{fig:rep},
for the case $j_0=3/2$.) 

We conclude that the Hilbert space, in the representation with $(N^2, N\cdot J) = (1, s\hbar)$, is indeed spanned by the states defined in (\ref{basis}), with $j_0 = |s|$. As a consequence, we see that the Hilbert space is non-trivial if and only if 
\beq\label{sinz/2}
s\in \frac{1}{2}\bb Z\, ,
\eeq
since $j$ can only take values in $\frac{1}{2}\bb Z^{0+}$.

\subsection{Dirac's condition and intrinsic spin}
\label{ssec:DiracC}

According to the matching condition for the Casimir \eqref{CCasimir}, and the definition of $s$ \eqref{NJs},
the quantum theory is based on the representation with $s = - eg/\hbar$. The 
requirement for the theory to
be non-trivial, 
\eqref{sinz/2}, 
thus implies
\begin{equation}\label{monoindex}
eg = \frac{n}{2}\hbar \, , \quad n \in \bb Z  \,,
\end{equation}
which is precisely Dirac's charge quantization condition. 

It is interesting to note how the more restrictive Schwinger condition \cite{schwinger1966magnetic, schwinger1975magnetic,peres1968rotational}, $eg = n\hbar$, would appear in this approach. In the previous section we derived the constraint on $s$ by analyzing the representations of the algebra generated by $J$'s and $N$'s. However, not all of these can be ``integrated'' to representations of the group $G = \bb R^{3*} \rtimes SO(3)$. As mentioned before, the representations of the algebra are in correspondence with representations of the universal cover of the group, $\widetilde G = \bb R^{3*} \rtimes SU(2)$. 
In order to have a true representation of $SO(3)$, rather than just a projective representation
(i.e.\ a representation up to a phase),
one must impose that a rotation by $2\pi$ corresponds to the identity operator, which 
implies that only integer spins, $j \in \bb Z^{0+}$, are allowed. 
Were one to insist that the quantum theory be based on true representations of the quantizing group, $G$, 
this would thus impose the constraint $s \in  \bb Z$, which leads to Schwinger's condition. 
However, there is no fundamental reason within quantum mechanics to exclude projective representations 
as realizations of a symmetry.

Let us now make a comment about intrinsic spin. In addition to the interpretation given in Section \ref{ssec:CasInv} for the Casimir $N \cdot J$, as $-eg$, there is another interpretation, as a measure of the {\sl intrinsic spin} of the particle. To see this, consider a basis of simultaneous eigenvectors of $N = (N_1, N_2, N_3)$, denoted by $|n_1, n_2, n_3\rangle$. In the representation with $N^2 = 1$ we must have $n_1^2 + n_2^2 + n_3^2 = 1$. This can be interpreted as a (non-normalizable) state localized at the point $n = (n_1, n_2, n_3)$ on the sphere. Without loss of generality, consider such a state localized at the north pole $|u\rangle = |0,0,1\rangle$. We have
\begin{equation}\label{u-state}
J_3 |u\rangle = J\cdot N |u\rangle  = s\hbar  |u\rangle  \,,
\end{equation}
 so that $|u\rangle$ is an eigenstate of $J_3$ with eigenvalue $s\in\frac12 \bb Z$.
 But $|u\rangle$ is localized at the north pole, so this angular momentum must be an intrinsic spin of the particle. Hence, there is an equivalence between a spinless particle with electric charge $e$ in the presence of a magnetic monopole of charge $g$ inside the sphere, and a particle with spin $s\hbar = -eg$, with no magnetic monopole \cite{fradkin2013field, kikuchi2015spin}. Interestingly, the same equivalence occurs in classical physics:  a free particle with spin would precess on a circle smaller than the great circle, conserving angular momentum, exactly as if there were a magnetic field curving the orbit of a spinless charged particle. In fact, if it is not possible to turn off the magnetic monopole (or modify its charge), and the particle is truly living on a sphere (without access to higher dimensions), then there is no way to distinguish a magnetic monopole from an intrinsic spin.

\subsection{Energy spectrum}
\label{ssec:Energy}

In this section we compute the energy spectrum for a non-relativistic particle of mass $m$ living on a round 
sphere
of radius $r$, in the presence of a magnetic field given by (\ref{B}). We assume here that the magnetic field is uniform with respect to the metric on the sphere.  
The classical time-evolution Hamiltonian is
\begin{equation}\label{H}
H = \frac{1}{2m}h^{ab}p_ap_b  \,,
\end{equation}
where $p_a$ is the canonical momentum---a cotangent vector on the sphere---and 
$h_{ab}$ is the metric on the sphere (using abstract tensor index notation).
Note that the magnetic field does not appear in the Hamiltonian for, in our approach, it is fully encoded in the symplectic form. 
From Hamilton's equations, we see that the kinematical velocity is simply related to the momentum by 
\begin{equation}
\dot x^\a = \{x^\a, H\}  = \frac{1}{m}  h^{\a\b} p_\b \,,
\end{equation}
where $x^\a = (x^1, x^2)$ are any coordinates for the sphere.  
Thus, on a solution,
the value of the Hamiltonian is the kinetic energy, $\frac12 m h_{\a\b}\dot x^\a\dot x^\b$. 

Up to this point only the topology of the sphere, and the choice of an $SO(3)$ subgroup of the diffeomorphisms preserving $B$, has played a role in
identifying and quantizing the canonical group. Now we identify this subgroup with the symmetry group of the metric $h_{ab}$, and use that identification to express the Hamiltonian in terms of the canonical observables
$J$ and $N$. To this end, we first express the inverse metric of the sphere in terms of the rotation Killing vector fields $\overline{X}_i := \overline{X}_{n = e_i}$  (cf.\ footnote \ref{fn12})
 and the geometrical radius $r$,
 \beq\label{h}
 h^{ab}= \frac{1}{r^2}\, \overline{X}_i^a\, \overline{X}_i^b \,,
 \eeq
with implicit summation
over the repeated index $i$.\footnote{To verify that this sum yields the inverse metric, note first that it is invariant under rotations. The normalization constant can be determined by looking at a specific point, namely $(\theta, \phi) = (\pi/2, 0)$. There we have $\overline{X}_1 = 0$, $\overline{X}_2 = \partial_\theta$ and $\overline{X}_3 = \partial_\phi$,
and the inverse metric thus reads $r^{-2}(\partial_\theta^2 + \partial_\phi^2)$, as desired.}
Inserting \eqref{h} for $h^{ab}$ in the Hamiltonian \eqref{H} yields
\begin{align}
H &= \frac{1}{2mr^2}(\overline{X}_i^a p_a)(\overline{X}_i^bp_b)\nonumber\\
&= \frac{1}{2mr^2}(J_i + eg N_i)(J_i + eg N_i)\nonumber\\
&=  \frac{1}{2mr^2}[J^2- (eg)^2]\, , 
\end{align}
where we used \eqref{Pn1}, \eqref{Qa1}, and \eqref{JiNidef} in the second line, 
and \eqref{CCasimir} in the third line.
To quantize, we just replace the function $J$ by the quantum operator $J$. 
The Hamiltonian is clearly diagonal in the basis $|j,m\rangle$, and the energy eigenvalues are
defined by 
\begin{equation}
H|j,m\rangle =  \frac{\hbar^2}{2mr^2} \left[ j(j+1) - j_0^2 \right]|j,m\rangle  \,,
\end{equation}
where $j_0 = |s|=|eg|/\hbar$, and $j-j_0\in \bb Z^{0+}$. 
Equivalently, in terms of the  non-negative integer 
$k = j - j_0$, 
the energies can be enumerated as
\begin{equation}
E_{km} =  \frac{\hbar^2}{2mr^2} \left[ k(k+1) + |s| \left(2k+1 \right) \right]  \,.
\end{equation}
The degeneracy of these ``Landau levels'' 
arises only from $m$, so there are $2j+1 = 2(k + |s|) + 1$ states 
at level $j$. 
In particular, the ground state has degeneracy $2|s|+1$, 
corresponding to one additional state for each magnetic flux quantum.
This agrees with other derivations of the energy spectrum \cite{kemp2014geometric, wu1976dirac, haldane1983fractional, greiter2011landau}.

The case where the magnetic field is not uniform with respect to the metric is slightly more subtle. In principle, we could include it in the symplectic form and proceed as before.
However, unless $B$ is uniform, no $SO(3)$ subgroup of the $B$-preserving diffeomorphisms of the sphere would be an isometry of the metric. Consequently, there would be no preferred form for the Hamiltonian when written in terms of the canonical observables, 
leading to operator-ordering ambiguities in the quantization. 
Instead we can split the magnetic field into its monopole and higher multipole parts, continuing to include the monopole part in the symplectic structure, but incorporating the higher multipole part into the Hamiltonian via the usual minimal coupling. By doing so,  we obtain a globally defined Hamiltonian,
which admits a standard application of canonical quantization.
We discuss this case in Appendix \ref{app:ih}, where we also address the issue of gauge invariance in Isham's group theoretic quantization scheme.

\section{Discussion}
\label{sec:Dis}

In this paper we have studied the problem of quantizing a particle on a 2-sphere in the presence of a magnetic monopole using Isham's group-theoretic scheme. Based on the principles of canonical quantization, the quantum theory is constructed from unitary irreducible (projective) representations of a transitive group of symplectic symmetries of the phase space. Our goal was to analyze the problem in a rigorous manner,
emphasising the role of Casimir invariants in connecting the classical and the quantum worlds. To ensure robustness and generality of the quantization, we referred only to intrinsic properties of the system, adopted a gauge-invariant approach, and did not make any a priori assumptions about the Hilbert space.

In order to formulate the problem in a gauge-invariant language, with globally defined objects, we described the magnetic monopole as a ``flux term'' in the symplectic form on the phase space. 
A natural transitive group of symmetries of the phase space is the Euclidean group, $E_3 = \bb R^{3*} \rtimes SO(3)$, where $SO(3)$ implements spatial rotations of the sphere and $\bb R^{3*}$ corresponds to momentum translations (i.e., boosts). 
The canonical observables consist of three angular momentum ``coordinates'', $J$,
generating the spatial rotations, and three position ``coordinates'', $N$, generating the momentum translations.
Although the group structure is independent of the magnetic monopole, the presence of the latter affects the expressions of the canonical observables, $J$ and $N$, as functions on the phase space, {and it also affects the value of the classical Casimir 
$J\cdot N = -eg$.}
To construct the Hilbert space, we employed an algebraic method involving ladder operators that raise and lower the eigenvalue of the $J^2$ operator. This method for deriving the unitary irreducible representations of the Euclidean group resembles the usual approach for $SU(2)$. By imposing that the theory is free of negative-norm states, one gets a constraint on the possible values of the Casimir invariants $N^2$ and $J \cdot N$. If we set $N^2 = 1$ (which can always be done without loss of generality), 
then $J\cdot N$ must be an integer multiple of $\hbar/2$. Insisting that the values of these quantum invariants must be the same as their classical counterparts, we obtain Dirac's charge quantization condition, $eg=n\hbar/2$.

Although Isham's scheme provides a well-founded algorithm for quantization, it still suffers 
from ambiguities.
One ambiguity appears in the choice of the particular $SO(3)$ subgroup of the (infinite-dimensional) group of volume-preserving diffeomorphisms of the sphere, whose lift to the phase space was used to construct the quantizing group. Although we argued that this choice does not affect the Hilbert space, it is relevant for the dynamics of the quantum theory. Indeed, the form of the Hamiltonian as a function of the canonical observables depends on which $SO(3)$ subgroup is chosen. For a generic metric, no such a choice is preferred and thus the Hamiltonian would have no preferred form, leading to operator-ordering ambiguities.
We contrast this with the case of the round sphere, where the group of isometries of the metric provides a preferred $SO(3)$ subgroup of symmetries, and leads to a simple (free) Hamiltonian that can be quantized unambiguously. Save for a few other highly symmetric cases (like the ellipsoid), some additional principle would be needed to resolve the ambiguities associated with a generic geometry.  
Another, more fundamental, ambiguity that is potentially resolved by the metric is the choice of the quantizing group itself. In fact, there are many finite-dimensional groups that act transitively on the sphere (e.g., the Lorentz group), so it may be that the underlying justification for the choice of $SO(3)$ comes down to the symmetries of the metric. 
{However the example of a particle on $S^3$ reveals that the metric may not be sufficient to cure this ambiguity, as in this case both $SU(2)$ and $SO(4)$ act transitively and isometrically on a round sphere. For the case of a particle on $S^3$, see for example \cite{aldaya2008group, aldaya2008quantum, aldaya20162, guerrero2020particle}.}

It is important to stress that this procedure is predicated on the assumption that the classical theory, and in particular all of the observables, are defined globally on phase space, since the quantum description is inherently global.
To illustrate this point, 
suppose that, instead of incorporating the magnetic monopole field in the 
symplectic form, we took the more common route of using the canonical sympletic form and including the 
vector potential $A$ in the Hamiltonian, 
$H = \frac{1}{2m}(p - eA)^2$.
It would still be natural to choose the Euclidean group 
for the quantizing group. If we did so,
the Casimir $J\cdot N$ would vanish classically
and, if the Casimir matching principle were to apply, this would mean that 
 $J\cdot N$ would also vanish quantum mechanically. But 
 the quantum theory should not depend on the particular way one decides to describe the classical system, and we have already established that 
 in the quantum theory $J\cdot N$ should be equal to $-eg$.
This failure of the Casimir matching principle can be attributed to 
the failure of the classical phase space description to be global. 
Since no globally defined vector potential $A$ can describe the monopole, 
one must cover the sphere in at least two overlapping gauge patches. Classically, one can focus on phase space trajectories that are temporarily confined to one or the other patch, switching descriptions in the overlap region, but that precludes a global map between the classical and quantum observables. This is not to say that one cannot describe the quantum theory using the vector potential for the monopole, but only
that it cannot be done according to the standard framework of global canonical quantization.
Instead, one can ensure that the quantum description itself is globally well defined.
As explained in Appendix \ref{app:Bundle}, in such a global description the wave functions 
are sections of a complex line bundle over the sphere 
carrying a representation of the Euclidean group.
In such a representation, 
$p - eA$ can be realized as a covariant derivative 
provided that the product of the charges satisfies the Dirac quantization 
condition $eg=n/2$ for some integer $n$, and that
the Chern number of the bundle is $n$. In this formulation,
the quantum Casimir $J\cdot N$ takes the correct value, $-eg$.

As the previous paragraph makes clear, 
when quantizing a classical system
the determination of the correct Hilbert space 
can depend not only on the 
intrinsic structure of the phase space, but on the 
dynamics as well. For the particle on the sphere in a 
monopole field, the impact of the dynamics on the 
Hilbert space is felt either from the Casimir matching principle (when the dynamical effects of the magnetic monopole are encoded in the symplectic form), or from the Chern class selection (when the monopole is encoded in the Hamiltonian via the gauge potential). This role of the dynamics is absent in quantum mechanics on $\bb R^n$ since 
(by the Stone-von Neumann theorem)
the canonical algebra of 
position and momentum coordinates
has a unique representation, so the Hamiltonian plays no role in selecting the Hilbert space. On the other hand, it is ubiquitous in quantum field theory,
where the infinite dimensionality of the algebra allows for many representations of the canonical algebra, and the selection of a representation depends on other physical observables, such as the Hamiltonian. 
For example, as shown by Haag's theorem~\cite{Coester:1960lxe,haag2012local}, the representations containing a translation invariant vacuum state in infinite volume differ even for a free scalar field  with different masses~\cite{duncan2012conceptual}. 
The problem of a particle on a sphere provides a simple example where the non-uniqueness of the representation, resolved only by the dynamics, comes from the non-trivial topology of the phase space, rather than from the infinite dimensionality of the algebra.

The difference between the quantizations on a plane and on a sphere can be understood from a group-theoretic perspective, in terms of the ``planar limit'' of the sphere, as follows.
We expect that a particle that remains near the north pole of the sphere at all times should not be able to ``feel'' the global structure of the sphere. Thus, in some limit, the quantum mechanics on a sphere must 
reduce to
the usual one on a plane. To see how this works, consider a ``sector'' of the Hilbert space in which $X_1 := N_1 \sim \theta$, $X_2 := N_2 \sim \theta$ and $N_3 \approx 1$, where $\theta \ll 1$ is the angle around the north pole where the particle is localized.
Note that the ``vertical momentum'' $P_3 \sim \theta|P|$ is small in this sector, so we can approximate $J_1 \approx - P_2 - eg X_1$, $J_2 \approx P_1 - eg X_2$ and $J_3 \approx X_1 P_2 - X_2 P_1 - eg$, where terms of order $\theta^2$ were neglected. In this sector, the algebra reduces to $[X_i, P_j] = i\hbar \delta_{ij} N_3$, $J_3$ behaves as the generator of rotations for $X_i$ and $P_i$, and $N_3$ becomes a central element (taking the value $1$ in the relevant
representation). This deformation of the algebra is known as the In\"on\"u-Wigner contraction. At the group level, this corresponds to a deformation of the Euclidean group $E_3 = \bb R^3 \rtimes SO(3)$ into $(\bb R^2 \times \bb R) \rtimes (\bb R^2 \rtimes SO(2) )$, where the first factor is generated by $(X_1, X_2; N_3)$ and the second by $(P_1, P_2; J_3)$. This contracted group can be reexpressed as $H(2) \rtimes SO(2)$, where $H(2)$ is the Heisenberg group in two spatial dimensions, generated by $(X_1, P_1, X_2, P_2; N_3)$, and $SO(2)$ is the rotation group around the origin, generated by $J_3$. Note that $S := J_3 - (X_1 P_2 - X_2 P_1) = -eg$ is a Casimir operator, interpreted as the intrinsic spin of the particle. 
Since $J_3$ 
differs from 
$X_1 P_2 - X_2 P_1$ 
only by a Casimir operator,
it follows that 
the irreducible representations of $H(2) \rtimes SO(2)$ are also irreducible when restricted to $H(2)$. 
As
$H(2)$ has a unique irreducible unitary representation,
this
confirms that we do in fact recover the quantum mechanics on a plane (for {\it any} value of the intrinsic spin Casimir $S$).
It is interesting to note an important difference between the plane and the sphere: the subgroup $SO(2)$ of $E_3$ was ``pulled out'' of $SO(3)$ during the deformation, so it appears as a factor in $H(2) \rtimes SO(2)$
rather than as a subgroup of $SO(3)$. 
The spin is therefore {\it not} quantized on a plane,
because the $SO(2)$ gets ``unwrapped'' to $\bb R$ when considering projective representations (that is, when considering the universal cover of the group). Thus the quantization of the spin on a sphere is a truly topological effect. 

In conclusion, we have seen that the problem of quantizing a particle on a sphere is quite effective in revealing some of the subtleties associated with a non-trivial phase space topology. It also serves to illustrate how Isham's scheme, which provides a general class of quantum theories compatible with the classical kinematics, must be paired with additional principles (e.g., Casimir matching) and considerations about dynamics in order to single out a preferred quantum theory. 
{The fact that non-trivial phase space topology affects global aspects of quantization
applies as well to quantum field theories, for which the implications are not
yet fully known.}
In particular,
the longstanding problem of quantizing general relativity, as a candidate theory of quantum gravity, is a prime example: 
even before imposing the  Hamiltonian constraints, {if the configuration space is given by metrics (or co-frame fields) on a spatial slice,  the condition that these are non-degenerate (and of positive signature) leads to a phase space with  non-trivial topology}. This suggests that standard
canonical commutation relations are not appropriate, and that instead the quantization should be based on an affine 
algebra~\cite{Klauder:1970ut,
pilati1982strong, pilati1983strong,
 klauder1999noncanonical, klauder2002affine},
 a conclusion that is also 
 reached by the application of 
 Isham's global, group theoretic quantization scheme~\cite{isham1984groupI, isham1984groupII}.

\section*{Acknowledgements}
We are grateful to  M.~Greiter for helpful correspondence. 
This research was supported in part by the National Science Foundation under Grants PHY-1708139 at UMD and PHY-1748958 at KITP.

\appendix

\section{No central extensions for the $\ca E_3$ algebra}
\label{app:Euc}

In this section we offer a simple derivation of the fact that the  algebra $\ca E_3$ 
\eqref{algJN}
of the Euclidean group $E_3$ 
does not admit non-trivial central extensions by 2-cocycles. {Although this result is presumably well-known, we have not found a reference proving it explicitly.}\footnote{{There are some general theorems that can be applied to the Euclidean algebra, such as Proposition 1 of \cite{vizman2004central} or Proposition 14.1 of \cite{vandenban2004rep}.
These theorems imply that $\ca E_n$ 
does not admit non-trivial central extensions except for $n=2$.}}

Let $\al g$ be a Lie algebra and let $\widetilde{\al g} = \al g \oplus_S \bb R$ be a central extension. 
If $\xi \in \al g$ and $r \in \bb R$, we generically define the product on $\widetilde{\al g}$ as
\begin{equation}
\left[(\xi, r), (\xi', r')\right] = \left([\xi, \xi'], \theta(\xi, \xi') \right) \,,
\end{equation}
for some function $\theta : \al g^2 \rightarrow \bb R$. If this is to form a Lie algebra, $\theta$ must be linear, antisymmetric and, due to Jacobi identity, satisfy
\begin{equation}
\theta(\xi, [\xi', \xi'']) + \theta(\xi', [\xi'', \xi]) + \theta(\xi'', [\xi, \xi']) = 0 \,,
\end{equation}
which is called the 2-cocycle condition. The extension is said to be trivial if $\theta(\xi, \xi') = f([\xi, \xi'])$ for some linear $f : \al g \rightarrow \bb R$. (In this case, note that $\xi \mapsto (\xi, f(\xi))$ is a homomorphism from $\al g$ to $\widetilde{\al g}$.)

Consider now the $\ca E_3$ algebra. Define
\begin{align}
&\theta_{ij} = \theta(J_i, J_j) \nonumber\\
&\theta_{\alpha\beta} = \theta(N_\alpha, N_\beta) \nonumber\\
&\theta_{\alpha i} = \theta(N_\alpha, J_i) \,,
\end{align}
where Latin and Greek indices are used to distinguish these components (i.e., $\theta_{ij}$ represents a different set of numbers than $\theta_{\alpha\beta}$). These components can be rewritten as
\begin{align}
&\theta_{ij} = \epsilon_{ijk} w_k \nonumber\\
&\theta_{\alpha\beta} = \epsilon_{\alpha\beta\gamma} h_\gamma \nonumber\\
&\theta_{\alpha i} = \epsilon_{\alpha i \beta} w_\beta  \,,
\end{align}
for the nine numbers $w_i$, $h_\alpha$ and $w_\alpha$.

For a trivial extension, we must have
\begin{align}
&\theta_{ij} = \theta(J_i, J_j) = f([J_i, J_j]) = f( \epsilon_{ijk} J_k) = \epsilon_{ijk} f_k \nonumber\\
&\theta_{\alpha\beta} = \theta(N_\alpha, N_\beta) = f([N_\alpha, N_\beta]) = 0 \nonumber\\
&\theta_{\alpha i} = \theta(N_\alpha, J_i) = f([N_\alpha, J_i]) = f(\epsilon_{\alpha i \beta} N_\beta) = \epsilon_{\alpha i \beta} f_\beta \,,
\end{align}
where $f_i = f(J_i)$ and $f_\alpha = f(N_\alpha)$. We conclude that the extension is trivial if and only if $h_\alpha = 0$, for in this case we can always define $f_i = w_i$ and $f_\alpha = w_\alpha$.

Consider the cocycle condition for two $N$'s and one $J$,
\begin{equation}
\theta(N_\alpha, [N_\beta, J_i]) + \theta(N_\beta, [J_i, N_\alpha]) + \theta(J_i, [N_\alpha, N_\beta]) = 0 \,,
\end{equation}
which gives $\epsilon_{\beta i \gamma} \theta_{\alpha\gamma} + \epsilon_{i\alpha\gamma} \theta_{\beta\gamma} = 0$. Or, in terms of $h_\alpha$,
$\delta_{i\alpha} h_\beta - \delta_{i\beta} h_\alpha = 0$. Contracting $i$ and $\alpha$, we get $h_\beta = 0$, proving that the $\ca E_3$ algebra admits only trivial central extensions.

\section{Details on the construction of the Hilbert space}
\label{app:Alg}

In this appendix we explain some of the details involved in the construction of the Hilbert space of the theory, presented in section \ref{ssec:Reps}. In particular, we want to derive equations (\ref{N3})-(\ref{N-j0}). For simplicity, we use $\hbar = 1$ in this section.

Note first that the angular momentum algebra, 
\beq
[J_3,J_\pm]=\pm J_\pm \,,\qquad [J_+,J_-]=2J_3 \,,
\eeq
alone yields the action of $J_i$ on the basis states
\eqref{basis}: 
\begin{align}
J_3|j,m\ra&=m|j,m\ra \,,\\
 J_-|j,m\ra&=|j,m-1\ra \,,\\ 
 J_+|j,m\ra&=\left[ j(j+1)-m(m+1) \right]|j,m+1\ra \,.\label{J+jm}
 \end{align}
Moreover, the norms of all the states $|j,m\ra$ are related to the norm of $|j,j\ra$ recursively, via 
\begin{align}
\la j,&m-1|j,m-1\ra = 
2m\la j,m|j,m\ra \,+ \nonumber\\
&+ [j(j+1)-m(m+1)]^2 \la j,m+1|j,m+1\ra \,,
\end{align}
hence they all have positive norm 
provided the edge states $|j,j\ra$ do. 
Thus we need only consider the action on and norms of the edge states.

For the rest, we use the algebra relations
\beq
[J_3,N_\pm]=\pm N_3 \,, \quad [J_\pm,N_3]=\mp N_\pm \,, \quad [J_\pm,N_\mp] = \pm 2 N_3 \,,
\eeq
(and  $[J_\pm,N_\pm] = 0$).
Using the last of these, the Casimir $N\cdot J = s$ can be written as
\begin{equation}
N\cdot J = N_3 (J_3 + 1) + \frac{J_- N_+ + N_- J_+}{2} \,,
\end{equation}
and applying this on $|j,j\ra$ yields
\begin{equation}\label{AppAN3}
N_3|j,j\rangle =  \frac{s}{j+1} |j,j\rangle  -\frac{1}{2(j+1)} |j+1,j\rangle  \,.
\end{equation}
To find $N_-|j,j\rangle$ for $j > j_0$, we can write 
\begin{align}\label{AppAN-}
N_- |j,j\rangle &= N_- N_+ |j-1,j-1\rangle\nonumber\\ 
&= \left( 1 - N_3^2 \right) |j-1,j-1\rangle  \,,
\end{align}
in which we set $N^2 = 1$. Using formula (\ref{AppAN3}) we have
\begin{align}
N_3^2& |j-1,j-1\rangle = N_3 \left( \frac{s}{j} |j-1,j-1\rangle - \frac{1}{2j} J_- |j,j\rangle \right) \nonumber\\
& = \frac{s}{j} N_3 |j-1,j-1\rangle - \frac{1}{2j} (J_- N_3 - N_-) |j,j\rangle  \,,
\end{align}
which together with \eqref{AppAN3} and  (\ref{AppAN-}) yields
\begin{align}
N_- |j,j\rangle &= \frac{2j}{2j+1} \left( 1 - \frac{s^2}{j^2} \right) |j-1,j-1\rangle \,+  \nonumber\\
+ \,&\frac{s}{j(j+1)} |j,j-1\rangle -\frac{|j+1,j-1\rangle}{2(2j+1)(j+1)}  \,, \label{N-app}
\end{align}
provided that $j>j_0$.

Next we compute the norms $\| N_+ |j,j\rangle\|^2$ and $\| L^{(j)} |j,j\rangle\|^2$, where $L^{(j)}$ 
is the $j$-lowering operator defined in (\ref{Lj}),
\begin{equation}\label{Lj2}
L^{(j)} = N_- - \frac{s}{j(j+1)} J_- + \frac{1}{2(2j+1)(j+1)} (J_-)^2 N_+ \,.
\end{equation}
In what follows, we use $\approx$ to denote operator identities that are valid only within $\langle j,j| \cdots |j,j\rangle$. For the raised state we compute
\begin{align}
&N_- N_+ = 1 - N_3^2 \approx \nonumber\\
&1 - \left( \frac{s}{j+1} -\frac{1}{2(j+1)} N_- J_+ \right) \left( \frac{s}{j+1} -\frac{1}{2(j+1)} J_- N_+ \right) \nonumber\\
& \qquad  \approx 1 - \frac{s^2}{(j+1)^2} - \frac{1}{2(j+1)} N_- N_+   \,,
\end{align}
and, solving for $N_- N_+$,
\begin{equation}
N_- N_+ \approx \frac{2(j+1)}{2j+3} \left( 1 - \frac{s^2}{(j+1)^2} \right)  \,,
\end{equation}
which yields \eqref{N-N+} for the squared norm.
For the lowered state we have
\begin{equation}
L^{(j)\dag} L^{(j)} \approx \frac{j(2j + 3)}{(j+1)(2j+1)} N_- N_+ - \frac{2s^2}{j(j+1)^2}  \,,
\end{equation}
and using the result for $N_- N_+$ we get
\begin{equation}
L^{(j)\dag} L^{(j)} \approx \frac{2j}{2j+1} \left( 1 - \frac{s^2}{j^2} \right)  \, ,
\end{equation}
which yields \eqref{LdL} for the squared norm.

\section{Wavefunctions and Chern numbers}
\label{app:Bundle}

In this appendix we review Mackey's approach, and apply it to the construction of the irreducible unitary representations of the Euclidean group $E_3$
(and of its universal cover), which  recovers the usual concept of wavefunctions as sections of complex vector bundles. 
We show in particular that the value of the Casimir 
$N\cdot J$, which is related to the magnetic charge, determines the first Chern number of the bundle 
via the Casimir matching requirement.
Interestingly, in geometric quantization \cite{woodhouse1997geometric, kemp2014geometric}, 
where quantum states are also given by sections of a line bundle,
the same relation between the magnetic charge and the bundle topology arises, albeit in a  different manner: by construction, the line bundle carries a connection whose curvature is required to coincide with the symplectic form \eqref{sym}. 
{Here we follow closely Isham’s presentation \cite{isham1984topological} of Mackey’s theory, using the language of fiber bundles. For a more technically complete presentation, framed in the language of measure theory, see for example \cite{mackey1969induced} or \cite{raczka1986theory}.}

As motivation, let us first consider only the $SO(3)$ part of the group. Given its natural action on $S^2$, $l_R(x) =: Rx$, 
we can construct a representation with wavefunctions $\psi : S^2 \rightarrow \bb C^d$ defined by
\begin{equation}\label{wavetriv}
\left( U(R)\psi \right) (x) = \psi(R^{-1}x)  \,,
\end{equation}
where $R \in SO(3)$ and $x \in S^2$. This representation is unitary with respect to the inner product $\langle \psi, \phi \rangle = \int\! d\mu\, \psi^* \phi$, where $d\mu$ is the Euclidean measure on the sphere, but it is not irreducible.\footnote{For example, in the case $d=1$, the space of complex functions on the sphere can be expanded in spherical harmonics, but since each $l$-subspace is invariant the representation is not irreducible.} Also, it is not exhaustive (i.e., not all unitary representations have this form). 

To generalize this, consider a Hermitian vector bundle over the sphere, $\bb C^d \rightarrow B \rightarrow S^2$.\footnote{In the notation $F  \text{(fiber)} \rightarrow E  \text{(total space)} \rightarrow M \text{(base space)}$, the second arrow represents the bundle projection map from the total space to the base space, 
while the first arrow merely indicates that each fiber of the bundle is an embedded copy of $F$.}
We want to take sections of this bundle, $\Psi : S^2 \rightarrow B$, as the vectors of the representation. Since the analogue of (\ref{wavetriv}) would involve a mapping between distinct fibers, we must require that the bundle admits a lift of the group action, $L_R : B \rightarrow B$, satisfying $\tau \circ L_R = l_R \circ \tau$ (compatibility with fiber structure) and $L_R \circ L_{R'} = L_{RR'}$ (compatibility with group structure). In that case, we define
\begin{equation}\label{wavelift}
\left( U(R)\Psi \right) (x) = L_R (\Psi(R^{-1}x))  \,.
\end{equation}
Note that this makes sense because $L_R$ maps the point $\Psi(R^{-1}x)$, on the fiber over $R^{-1}x$, to a point on the fiber over $x$. The Hermitian structure of $B$, with inner product on each fiber denoted by $(\, , )$, gives rise to an inner product on the space of sections defined by
\begin{equation}\label{innprodSO}
\langle \Psi, \Phi \rangle = \int_{S^2}\! d\mu(x)\, \left( \Psi(x), \Phi(x) \right)  \,.
\end{equation}
In order for the representation to be unitary with respect to this inner product, we must require that the group lift is compatible with the hermitian structure of the bundle, i.e., $\left( L_R z, L_R z' \right) = (z,z')$, where $z$ and $z'$ are points on the same fiber of $B$. Note that (\ref{wavetriv}) is the special case where both the bundle and the group lift are trivial. 

To find the most general wavefunction representation of $SO(3)$ on $S^2$, we must classify all the bundles $\bb C^d \rightarrow B \rightarrow S^2$ that admit such a lift of $SO(3)$. It is possible to show \cite{isham1989canonical} that any such a bundle must be associated to the ``master bundle'' $SO(2) \rightarrow SO(3) \rightarrow SO(3)/SO(2) \sim S^2$, with projection map equal to the quotient $q : SO(3) \rightarrow SO(3)/SO(2)$, 
via some homomorphism $\scr U : SO(2) \rightarrow U(d)$. More precisely, $B$ is necessarily isomorphic to $SO(3) \times_\scr{U} \bb C^d$, the bundle defined by the equivalence classes $[R,z] = [R h , \scr{U}(h^{-1})z]$, for $R \in SO(3)$, $z \in \bb C^d$ and $h \in SO(2) \subset SO(3)$, with projection map $q_{\scr U}([R,z]) = q(R)$. 
{The group action on this bundle  is given by $L_R[R',z]:= [RR', z]$.}
Note that $\scr{U}$ is a unitary representation of $SO(2)$, and we will later be interested in the irreducible ones. Since $SO(2)$ is abelian, its irreducible representations are one-dimensional,  so only the case $d = 1$ is relevant.
These representations are given by
\begin{equation}\label{scrU}
\scr{U}^{\!(n)}(\theta) = e^{-in\theta}  \,,
\end{equation}
where $n \in \bb Z$. 
The choice of this representation is the only discrete choice that enters the construction of the general representation of $E_3$ using Mackey theory, hence one can anticipate that $n$ must be related to the magnetic monopole index in (\ref{monoindex}).

We are now ready to consider the full group, $\bb R^{3*} \rtimes SO(3)$. For more transparency, let us consider first
 a generic group of the form $V \rtimes K$, where $V$ is a vector space and $K$ is a Lie group.\footnote{Mackey's theory also applies, with a minor modification, if $V$ is abelian and $K$ is a separable, locally compact group. When $V$ is not a vector space, we just need to replace below the dual space $V^*$ by the space of unitary characters $\text{Char}(V)$.} The product rule is given by
\beq
(v, k) (v', k') = (v + \rho_k v', kk') \,,
\eeq
where $v \in V$, $k \in K$ and $\rho : K \rightarrow \text{Aut}(V)$ is a left $K$-action on $V$. Later we shall particularize to $K = SO(3)$ and $V = \bb R^{3*}$, in the dual representation $\rho_R = l^*_{R^{-1}}$ (see (\ref{Glaw})). Since a generic element $(v, k)$ can be decomposed as $(v, e)(0, k)$,
where $e$ is the identity element of $K$, the operators representing $V \rtimes K$ on a Hilbert space $\ca H$ will factorize accordingly, $U(v, k) = U(v, e) U(0, k)$. We can define $A(v) := U(v, e)$ and $D(k) := U(0,k)$, so that
\beq\label{VD}
U(v, k) = A(v)D(k) \,.
\eeq
Thus, in classifying the representations of 
$V \rtimes K$, we can study the representations
of $V$ and $K$ separately, in the following manner.

Starting with $V$, define the self-adjoint generators $N(v)$ by
\beq
A(\lambda v) = e^{-i \lambda N(v)} \,,
\eeq
where $\lambda \in \bb R$. Since $V$ is abelian, we have  $[N(v),N(v')]=0$, and 
$N(v + \lambda v') = N(v) + \lambda N(v')$,
meaning that $N$ is a linear map from $V$ into a set of commuting, self-adjoint operators on $\ca H$. 
Accordingly, a simultaneous eigenvector $|\chi\ra$
of $N(v)$, for all $v$, determines an element $w \in V^*$, such that
\beq
N(v) |\chi\ra = w(v) |\chi\ra \,.
\eeq
Nothing requires the eigenvalues of $N(v)$ to be non-degenerate, so each $w$ may label a Hilbert (sub)space $\ca S_w \subset \ca H$. It follows from the group structure that the operator $D(k)$ maps $\ca S_w$ unitarily onto $\ca S_{\widetilde\rho_k w}$, where
$\widetilde\rho_k$ is the dual action of $K$ on $V^*$,
defined as $\widetilde\rho_k w(v) = w(\rho_{k^{-1}}v)$ for all $v\in V$. 
The Hilbert space $\ca H$ will be given by a ``direct sum'' (or rather, ``direct integral'') of $\ca S_w$ over $w$'s in some region of $V^*$. 
If $D(k)$ is to act in a closed fashion, such a region 
must consist of one or more 
orbits of $K$. To ensure irreducibility of the $V \rtimes K$ representation, we must take this region to be a {\it single} $K$-orbit $\ca O$ (or its closure) in $V^*$. Roughly speaking,
\beq
\ca H \sim  ``\oplus_{w \in {\ca O}}
\ca S_w" \,.
\eeq
More precisely, $\ca H$ will be the space of 
sections of a vector bundle over $\ca O$, with fibers $\ca S_w \sim \bb C^d$ (for some dimension $d$). To classify these representations
we must therefore classify the corresponding vector bundles.

{As explained before for the case $K = SO(3)$,} in order for a vector bundle
$\bb C^d\rightarrow B \rightarrow \ca O$
over $\ca O\sim K/H$ 
(where $H$ is the little group
corresponding to the orbit $\ca O$)
to carry a representation of $K$, it
must admit a lift of the $K$-action, and thus must be associated to the master bundle $H \rightarrow K \rightarrow K/H$ 
via a unitary irreducible representation $\scr U : H \rightarrow U(d)$ of $H$. 
Cross sections of this bundle, 
$\Psi: \ca O \rightarrow B=K \times_\scr{U} \bb C^d$,
form a linear space which 
 carries a representation of $V \rtimes K$. 
The element $(v,k)$ is represented by
\begin{equation}\label{waverep}
(U(v,k) \Psi)(w) =  e^{-i w(v)} \sqrt{\frac{d\mu_k}{d\mu}(w)} \, L_k \left(\Psi(\widetilde\rho_{k^{-1}}w) \right)  \,,
\end{equation}
where $w \in \ca O$ and $L_k$ is the lift of $K$ to $K \times_\scr{U} \bb C^d$ defined by $L_k [k', z] = [kk', z]$. The phase factor on the right-hand side is $A(v)$, and the rest is the factor $D(k)$, as in \eqref{VD}.
Note that $D(k)$ is analogous to \eqref{wavelift}, except for the Jacobian-like factor $d\mu_k/d\mu$, which deserves a few words.
In the case of $SO(3)$, it is possible to define the inner product \eqref{innprodSO} using the Euclidean measure on the spherical orbit, and the invariance of this measure under $SO(3)$ implies that $U(R)$ in \eqref{wavelift} is unitary, so this Jacobian-like factor is not needed.
In general, however, the orbit $\ca O$ may not admit an invariant measure,\footnote{In many cases, such as when $K$ is locally compact and $H$ is compact, the Haar measure on $K$ can be pushed down to $K/H$, defining an invariant measure on $\ca O$.} but fortunately it always admits a measure $\mu$ that is {\it quasi-invariant} under $K$. That is, $\mu$ and its push-forward $\mu_k := \widetilde\rho_{k*} \mu$ through $\widetilde\rho_{k}$ have the same sets of measure zero.\footnote{Given a measurable map $f: X \rightarrow Y$ and a measure $\mu$ on $X$, its push-forward to $Y$ is defined as $f_*\mu[B] = \mu[f^{-1}(B)]$, where $B$ is any Borel subset of $Y$ and $f^{-1}$ denotes the pre-image under $f$.} In order to make $U(v, k)$ unitary under such a measure, we must introduce the Jacobian-like factor ${d\mu_k}/{d\mu}$, called the {\it Radon-Nikodym} derivative of $\mu_k$ with respect to $\mu$, which is a positive ($\mu$-almost-everywhere) continuous function on $\ca O$ satisfying $\mu_k[B] = \int_B \frac{d\mu_k}{d\mu} d\mu$ for all Borel sets $B \subset \ca O$. Representations defined for equivalent measures (i.e., having the same sets of measure zero) are unitarily equivalent, and since there is only one quasi-invariant measure on $K/H$ (up to equivalence), the measure in \eqref{waverep} is determined by $\ca O$ (up to equivalence).

Note that these representations are labeled by the choice of the orbit $\ca O$ and the little group representation $\scr{U}$. 
These representations are irreducible as long as $\scr{U}$ is irreducible. If $V^*$ decomposes into regular $K$-orbits, meaning that there exists a Borel map $\zeta: V^*/K \rightarrow V^*$ that associates a dual vector to each orbit, then all unitary irreducible representations are generated in this way. This is Mackey's main result.

In the case of interest, $\bb R^{3*} \rtimes SO(3)$, the space where the orbits live is $(\bb R^{3*})^*$, which can be naturally identified with $\bb R^3$. Thus $\widetilde\rho_R$, which acts on $w \in \bb R^{3**}$ as $w \mapsto \rho^*_{R^{-1}}w =  l^{**}_{R}w$, acts on $x \in \bb R^3$ as $x \mapsto l_R x = Rx$. That is, $SO(3)$ acts on $(\bb R^{3*})^* \sim \bb R^3$ in just 
the standard way. The orbits $\ca O$ decompose into two classes: spheres (with any radius) and a point (at the origin). The little group for the first kind is $SO(2)$, while for the second it is $SO(3)$. Since the orbits are regular, the irreducible unitary representations are labeled by the radius $a \in \bb R^+ \cup \{0\}$ of the orbit and, for $a > 0$ (which is the case of interest), the integer $n\in \bb Z$ specifying the irreducible unitary representation $\scr{U}^{\!(n)}(\theta) = e^{-in\theta}$ of the little group $SO(2)$. For a given value of $n$, the Hilbert space  consists of sections of the line bundle 
$SO(3)\times_{\scr{U}^{\!(n)}}\bb C$.

How are the basic operators, $N$ and $J$, realized
on this Hilbert space? Since the Euclidean measure on $S^2$ is invariant under $SO(3)$, we do not have the Jacobian factor in (\ref{waverep}),
which simplifies to
\begin{equation}\label{waverep2}
(U(\alpha,R) \Psi)(x) =  e^{-i \alpha(x)/\hbar} L_R \Psi(R^{-1}x)   \,,
\end{equation}
Note that we have introduced an $\hbar$ in the phase factor, for notational convenience.
In analogy with (\ref{qexp}), we define more generally the generating operators $J_\eta$ and $J_\alpha$ via
\begin{align}
&U(\exp(0,\lambda \eta)) =: e^{-i \lambda J_\eta/\hbar} \nonumber\\
&U(\exp(\lambda \alpha,0)) =: e^{-i \lambda N_\alpha/\hbar}   \,,  \label{qexp2}
\end{align}
where $\eta \in \al{so}(3) \sim \bb R^3$ and $\alpha \in \bb R^{3*}$. It follows that
\begin{align}
&J_\eta = i\hbar \left. \frac{d}{d\lambda} U(\exp(0,\lambda \eta)) \right|_{\lambda=0} \nonumber\\
&N_\alpha = i\hbar \left. \frac{d}{d\lambda} U(\exp(\lambda \alpha,0)) \right|_{\lambda=0}   \,.  \label{qexpNJ}
\end{align}
Hence,
\begin{align}
&(J_\eta \Psi)(x) = -i\hbar \, \al D_\eta \Psi (x) \nonumber\\
&(N_\alpha \Psi)(x) = \alpha(x) \Psi(x)   \,, \label{qexpNJs}
\end{align}
where $\al D$ is a derivative operator defined by
\begin{equation}\label{derivop}
\al D_\eta \Psi (x) = - \left. \frac{d}{d\lambda}L_{R_{\lambda\eta}}\Psi\left(R_{\lambda\eta}^{-1}x\right) \right|_{\lambda=0}   \,.
\end{equation}
It satisfies $\al D_\eta(f\Psi) = f \al D_\eta \Psi + X_\eta (f) \Psi$, where $f: \ca O \rightarrow \bb C$ and $X_\eta$ is the vector field (tangent to $\ca O \sim S^2$) generated by $\eta$. 

We now wish to relate the Casimirs $N^2$ and $N\cdot J$ with the labels $a$ and $n$ of the wavefunction representations. In an orthonormal basis $e_i$ for $\bb R^3$, and dual basis $e^i$ for $\bb R^{3*}$, we have
$N^2= \sum_{i=1}^3 (N_{e^i})^2$, and 
$N_{e^i}\Psi(x) =e^i(x) \Psi(x)= x^i \Psi(x)$, so 
\beq
N^2 \Psi(x) = x^2 \Psi(x) = a^2 \Psi(x) \,.
\eeq
Thus, not surprisingly, $N^2$ corresponds to the radius squared, $a^2$, of the sphere. 
The representation scales trivially with $N^2$,
and the choice we made previously was $N^2=1$, so we here consider also the case $a=1$.

Next we consider $N\cdot J$, which as we now show is
related with the index $n$ of the bundle $SO(3) \times_{\scr U^{\!(n)}} \bb C$. We establish this in two independent ways. In the first way, we show how $N\cdot J$ is related to the little group representation \eqref{scrU}, which is labeled by $n$. In the second way, we show how $N\cdot J$ can be directly related to the integrated curvature of a suitably constructed connection, which directly yields the Chern number of the bundle, and thus the index $n$. In this second way, we will not need to invoke Schur's lemma, but rather it will be a consequence of the construction that $N\cdot J$ is proportional to the identity, similarly to what happened with $N^2$ above.

In the first way, we note that since $N\cdot J$ is a Casimir in the algebra, it must according to Schur's lemma be proportional to the identity, so it suffices to evaluate its action on a single state $\Psi(x)$ at any given point $x$. 
Let us take $x$ to be the north pole $u=(0,0,1)$, 
and construct the bundle $SO(3)\times_{\scr{U}^{\!(n)}}\bb C$ using the $SO(2)$ little group of $u$.
Then we have $N\cdot J \Psi(u) = J_3 \Psi(u)$ and, 
since $u$ is a fixed point of $R_{\lambda e_3}$, the value of $J_3 \Psi(u)$ depends only on the value of $\Psi$ at $u$ (as opposed to in a neighborhood of $u$).
Denoting $\Psi(u) = [1, z] \in SO(3)\times_{\scr{U}^{\!(n)}}\bb C$, where $1$ is the identity element of $SO(3)$ and $z \in \bb C$, we have
\begin{align}
J_3 \Psi(u) &= -i\hbar \, \al D_{e_3} \Psi(u) \nonumber\\
&= i\hbar \frac{d}{d\lambda}L_{R_{\lambda e_3}}\Psi(u)  \nonumber\\
&= i\hbar \frac{d}{d\lambda}[R_{\lambda e_3}, z] \nonumber\\
&= i\hbar \frac{d}{d\lambda}[1, \scr U(R_{\lambda e_3})z] \nonumber\\
&= i\hbar \frac{d}{d\lambda}e^{-in\lambda}[1, z] \nonumber\\
&= \hbar n \Psi(u) \,,
\end{align}
where $d/d\lambda$ is evaluated at $\lambda=0$ at every step. In the second line we used that $R^{-1}_{\lambda e_3}u = u$; in the third line we used the definition of the group lift to the associated bundle, $L_R[R', z] = [RR', z]$; in the fourth line we used the
defining property of the associated bundle $SO(3)\times_{\scr{U}^{\!(n)}}\bb C$,
$[R, z] = [Rh, \scr U(h^{-1}) z]$, with $h = R^{-1}_{\lambda e_3}$; and in the fifth line we used \eqref{scrU}. Therefore $J\cdot N\Psi(x) = \hbar n \Psi(x)$, so we conclude that the value of the Casimir invariant $J \cdot N$ is determined by the little group representation $\scr{U}^{\!(n)}$. Given the identification $J\cdot N = -eg$, 
we obtain the Schwinger condition $eg = -n\hbar$,
which is more restrictive than Dirac's condition (\ref{monoindex}).

In order to include also projective representations of the quantizing group, we must extend $SO(3)$ to $SU(2)$ and repeat the same analysis. The relevant master principal bundle is
then
the Hopf bundle $U(1) \rightarrow SU(2) \rightarrow SU(2)/U(1) \sim S^2$. Because the little group is  $U(1)$, the associated bundles are again constructed with
the representations $\scr U^{\!(n)}$, so the quantum states are represented by sections of $SU(2) \times_{\scr U^{\!(n)}} \bb C$. The group $SU(2)$ acts on the sphere as 
\beq 
e^{iv\cdot \sigma} (x \cdot \sigma) e^{-iv\cdot \sigma} = (R_{2v}x) \cdot \sigma \,,
\eeq 
where $v \in \bb R^3$, $x \in S^2 \subset \bb R^3$ and $\sigma = (\sigma_1, \sigma_2, \sigma_3)$ are the Pauli matrices. To obtain the usual normalization for the $\al{su}(2)$ algebra (i.e., with structure constants $f_{ijk} = \varepsilon_{ijk}$) we must take the basis $\{\sigma_1/2, \sigma_2/2, \sigma_3/2\}$, so that $e_3 = \sigma_3/2$, and thus $J_3 = J_{\eta = \sigma_3/2}$. Repeating the steps of the previous derivation, the only difference is that the little group phase factor is $\scr U(e^{i(\sigma_3/2)\lambda}) = e^{-in\lambda/2}$, which leads to $J_3\Psi(x) = \hbar n/2$. That is, $s = n/2$, matching Dirac's condition.

In the second way of evaluating $N\cdot J$, we note that the Chern number is a topological property of the bundle, as it does not
depend on the connection used to evaluate it, so we shall just
use the structures available to construct some arbitrary connection.
 The natural ingredient to use is the derivative operator $\al D$, which is defined in \eqref{derivop} in terms of the group lift $L_R$. A possible definition for a covariant derivative is
 \begin{equation}\label{covder}
\nabla_V \Psi (x) = \al D_{x \times V} \Psi (x)  \,,
\end{equation}
where $V \in T_x S^2 \subset T_x \bb R^3$. One can check that this $\nabla$ satisfies all properties of a covariant derivative, so it defines a connection on the bundle.
Using (\ref{qexpNJs}) we can write it as\footnote{From \eqref{Jivec} we see that $-i\hbar \nabla_V$ is nothing more than the ``natural'' quantization of the classical linear momentum along $V$, since $V \cdot p = - V \cdot (N \times J)$, recovering the picture that $p$ acts as a derivative on wave functions.}
\begin{equation}\label{connect}
\nabla_V \Psi = -\frac{i}{\hbar}\, V \cdot (N \times J) \Psi  \,.
\end{equation}

The curvature $F$ of this connection is defined by 
\begin{equation}
(\nabla_X \nabla_Y - \nabla_Y \nabla_X - \nabla_{[X,Y]}) \Psi = F(X,Y) \Psi  \,,
\end{equation}
for vector fields $X$ and $Y$ on $S^2$. Treating $X$ and $Y$ as vectors on $\bb R^3$, tangent to the sphere, and using (\ref{connect}), we obtain
\begin{equation}
F(X,Y) = - \frac{1}{\hbar^2} X^i Y^j [(N \times J)_i, (N \times J)_j ]   \,.
\end{equation}
Using the algebra \eqref{EucAlg} of $N$ and $J$ we get
\begin{equation}
F(X,Y) = \frac{1}{i \hbar} N^2 ( X \times Y ) \cdot J  \,.
\end{equation}
Since $X \times Y$ is normal to the sphere, it acts on wavefunctions (in the representation with $N^2 = 1$) as $\epsilon(X,Y) N$, where $\epsilon = \sin\theta d\theta \wedge d\phi$ is the area form on the unit sphere. Thus the curvature 2-form can be expressed as
\begin{equation}
F =  \frac{1}{i \hbar}\, \epsilon\, N \cdot J \,.
\end{equation}
We see that $N\cdot J$ must act on sections as a function, i.e., $N\cdot J\, \Psi(x) = \hbar s(x) \Psi(x)$, for some $s:S^2 \rightarrow \bb R$. Because the connection was constructed in a rotation invariant way, the curvature must also be rotation invariant. This implies that $s$ is actually a constant, which is consistent with the fact that $N\cdot J$ is a Casimir operator in an irreducible representation.
The corresponding first Chern number is then
\begin{equation}\label{C1}
C_1 = \int_{S^2} \frac{i}{2\pi} F = \frac{s}{2\pi} \int_{S^2}  \epsilon = {2s}\,,
\end{equation}
which shows that the Casimir $N \cdot J$ is directly related to this topological number, as anticipated. 
The possible values of $s$ are quantized,
since the bundle $SO(3) \times_{\scr U^{\!(n)}} \bb C$ has first Chern number $2n$, while the extended bundle $SU(2) \times_{\scr U^{\!(n)}} \bb C$ has first Chern number $n$.
Note that this is consistent with our previous result, where $s=N\cdot J/\hbar$ was shown to be related to the bundle index $n$.

\section{Non-uniform magnetic fields}
\label{app:ih}

In this appendix we discuss the case where the magnetic field is not uniform with respect to the metric on the sphere. In principle, we could proceed as before, including the magnetic field in the symplectic form, so maintaining gauge invariance explicitly. However, unless $B$ is uniform, no $SO(3)$ subgroup of the $B$-preserving diffeomorphisms of the sphere would be an isometry of the (round) metric. Consequently, there would be no preferred form for the Hamiltonian, leading to operator-ordering ambiguities in the quantization. 
{Thus, it is convenient} to split $B$ into its monopole and higher-pole parts, 
$B = g\epsilon + dA$, where $\epsilon$ is the normalized volume form (i.e., whose integral is $4\pi$) invariant under the isometries of the metric and $A$ is a globally-defined potential 1-form, and then include only the monopole term, $g\epsilon$, in the symplectic form, as in \eqref{sym}, while including the {\it higher-pole} part, $dA$, in the Hamiltonian via the usual minimal coupling. In this way, the quantizing group and the associated canonical observables, $J$ and $N$, depend only on the monopole term, while the higher-pole term affects only the Hamiltonian.

In a given global gauge  
the Hamiltonian reads
\beq
H = \frac{1}{2m} (p - e A)^2\,,
\eeq
in which $(p - eA)^2 = h^{ab} (p - e A)_a (p - e A)_b$, where $h$ is the metric on the sphere.
Assuming a round metric, and using expression \eqref{h} in terms of the Killing vector fields $X_i$ (generated by $J_i$), the Hamiltonian becomes
\beq
H = \frac{1}{2mr^2} \left[
J_i - eg N_i - e A(X_i) \right]^2 \,,
\eeq
where a summation over $i \in \{1, 2, 3\}$ is implicit. 
Since $A(X_i)$ is a function of the position only, it can be written in terms of $N$ unambiguously. 
In the wavefunction realization of Appendix \ref{app:Bundle}, it acts simply by multiplication. 

Now we must worry about gauge-invariance. In particular, we must ensure that the same theory is obtained if one uses another choice of (global) gauge $A'$.
Since the sphere is simply connected, we have
\beq
A' = A + d\sigma \,,
\eeq
for some function $\sigma : S^2 \rightarrow \bb R$. Note that
\beq
H' = \frac{1}{2m r^2} (p - e A')^2 = \frac{1}{2m r^2} (p - ed\sigma - e A)^2 \,,
\eeq
which has the same form as the original Hamiltonian if we define a new momentum variable $p' = p - d(e\sigma)$. This corresponds to a momentum translation, defined as
\beq\label{pshift}
K_\sigma(p) = p - d(e\sigma) \,,
\eeq
which is a symplectomorphism of the phase space satisfying $H' = K_\sigma^* H$. Note that it maps the {symplectic} flow of $H'$ into that of $H$ and, being vertical on the phase space, it leaves unchanged the projection of the dynamical trajectories to the configuration space. Therefore the two gauge-related Hamiltonians produce equivalent classical dynamics. 
As to the quantization,
note first that $H' = K_\sigma^* H$ implies that $H'$ has the same functional form when written in terms of transformed charges $Q' = K_\sigma^*Q$ as $H$ written in terms of $Q$. That is, if $H = f(Q)$ then $H' = f(Q')$. 
Since the Poisson brackets is defined from the symplectic structure, which is invariant under $K_\sigma$, we have that the algebra of charges is preserved under such a transformation, i.e., $ \{K_\sigma^*Q_i, K_\sigma^*Q_j\} = K_\sigma^*\{Q_i, Q_j\}$. 
Thus the charges $Q'$ satisfy the same algebra as $Q$. Moreover the Casimirs are functionals of the charges, with form depending only on the algebra,
so it should be that $C' = K^* C$. But since $C$ is constant on the phase space, we have $C' = C$. Consequently, as the charges satisfy the same algebra, with the same Casimir values, the quantizations are equivalent 
(provided the same ordering prescription is applied to the Hamiltonian).

At the group level, the charges $Q$ and $Q'$ generate 
the same group $G$, but realized differently on the phase space. Namely, if the charges $Q$ generate
a realization $\Lambda : G \rightarrow \text{Diff}(\ca P)$ of $G$ as symplectomorphisms of $\ca P$, then it can be shown that $Q'$ generate
the transformed realization, $\Lambda'$, defined by
\beq\label{ClassConj}
\Lambda'_g = K_\sigma^{-1} \circ \Lambda_g \circ K_\sigma \,.
\eeq
Therefore, we see that a change of gauge in the Hamiltonian can be ``reversed'' by simply changing the way that the quantizing group 
acts on the phase space. 
Since it is the same group (with the same Casimir values) which is undergoing quantization, the same quantum theory is obtained.
At the quantum level, this change of realization corresponds to a unitary transformation on the Hilbert space. Mirroring \eqref{ClassConj}, the $K$-transformation is implemented as
\beq
U(\Lambda'_g) = T(K_\sigma)^\dag U(\Lambda_g)T(K_\sigma) \,,
\eeq
where $T(K_\sigma)$ is a unitary transformation, ensuring that the representation
 $U(\Lambda'_g)$ is equivalent to $U(\Lambda_g)$.
Since $K_\sigma$ is simply a generalized type of momentum translation, the action of $T$ is  defined, on the wave functions of Appendix \ref{app:Bundle}, analogously to $N$ in \eqref{qexpNJs}, as 
\beq
T(K_\sigma) \Psi(x) = e^{-ie\sigma(x)/\hbar}\Psi(x) \,.
\eeq
To verify that this is the desired transformation, define the ``momentum operator'' by
\beq
p(V) = V \cdot ( - N \times J) = - i\hbar \nabla_V  \,,
\eeq
where $V$ is a vector field on the sphere and $\nabla$ is the covariant derivative defined in \eqref{covder}. Conjugating by $T(K)$ we get
\beq
T(K_\sigma)^\dag p(V) T(K_\sigma) \Psi(x)  = (p - ed\sigma)(V) \Psi(x) = p'(V) \Psi(x) \,,
\eeq
where $p'$ is precisely the momentum operator defined from the modified charges, i.e., $p'(V) =  V \cdot( - N' \times J')$.
Therefore, since this implies that $H' = T(K_\sigma)^\dag H T(K_\sigma)$, the standard picture where a change of gauge corresponds to a phase transformation on wave functions is recovered.

\bibliographystyle{ieeetr}
\bibliography{BibRAS}

\end{document}